\documentclass[twocolumn,trackchanges]{aastex7}
\usepackage{graphicx}
\usepackage{subcaption}

\newlength{\toprowht}
\newlength{\botrowht}
\begin{document}

\title{Direct Measurement of Polar Coronal Hole-like Solar Wind in its Acceleration Phase}

\correspondingauthor{Samuel T. Badman}
\email{samuel.badman@cfa.harvard.edu}

\author[0000-0002-6145-436X]{Samuel T. Badman}
\affiliation{Center for Astrophysics $|$ Harvard $\&$ Smithsonian, Cambridge, MA, 02138, USA}
\email{samuel.badman@cfa.harvard.edu}


\author[0000-0002-8748-2123]{Yeimy J. Rivera}
\affiliation{Center for Astrophysics $|$ Harvard $\&$ Smithsonian, Cambridge, MA, 02138, USA}
\email{}

\author[0000-0002-7728-0085]{Michael Louis Stevens}
\affiliation{Center for Astrophysics $|$ Harvard $\&$ Smithsonian, Cambridge, MA, 02138, USA}
\email{mstevens@cfa.harvard.edu}  

\author[0000-0001-7401-5609]{Mari Paz Miralles}
\affiliation{Center for Astrophysics $|$ Harvard $\&$ Smithsonian, Cambridge, MA, 02138, USA}
\email{mmiralles@cfa.harvard.edu}  

\author[0000-0002-3699-3134]{Steven R. Cranmer}
\affiliation{Department of Astrophysical and Planetary Sciences,
Laboratory for Atmospheric and Space Physics,
University of Colorado, Boulder, CO 80309, USA}
\email{steven.cranmer@colorado.edu}

\author[0000-0003-4380-4837]{Andrea Verdini}
\affiliation{Dipartimento di Fisica e Astronomia, Università degli studi di Firenze, via Giovanni Sansone 1, 50019, Sesto Fiorentino, IT}
\email{andrea.verdini@unifi.it}

\author[0000-0002-4313-1970]{Lynn B. Wilson III}
\affiliation{NASA Goddard Space Flight Center, Heliophysics Science Division, Greenbelt, MD, USA.}
\email{lynn.b.wilsoniii@gmail.com}

\author[0000-0001-7019-5905]{Xiangyu Wu}
\affiliation{Mullard Space Science Laboratory, University College London, Holmbury St. Mary, Dorking, Surrey, RH5 6NT, UK}
\email{xiangyu.wu.23@ucl.ac.uk}

\author[0000-0003-3623-4928]{Georgios Nicolaou}
\affiliation{Mullard Space Science Laboratory, University College London, Holmbury St. Mary, Dorking, Surrey, RH5 6NT, UK}
\email{g.nicolaou@ucl.ac.uk}

\author[0000-0002-1628-0276]{Jean-Baptiste Dakeyo}
\affiliation{Space Sciences Laboratory, University of California, Berkeley, CA 94720-7450, USA}
\email{jbdakeyo@berkeley.edu} 

\author[0009-0008-6049-255X]{Kai Jaffarove}
\affiliation{Department of Earth, Planetary, and Space Sciences, University of California, Los Angeles, Los Angeles, CA 90095 USA}
\email{kaij@ucla.edu}

\author[0000-0002-8475-8606]{Tamar Ervin}
\affiliation{Physics Department, University of California, Berkeley, CA 94720-7300, USA}
\affiliation{Space Sciences Laboratory, University of California, Berkeley, CA 94720-7450, USA}
\email{tamarervin@berkeley.edu} 

\author[0000-0002-8650-1310]{Dominic S. Payne}
\affiliation{Climate and Space Sciences and Engineering, University of Michigan, Ann Arbor, MI 48109, USA}
\email{dspayne@umich.edu}

\author[0000-0003-4747-6252]{Michael Terres}\affiliation{Center for Astrophysics $|$ Harvard $\&$ Smithsonian, Cambridge, MA, 02138, USA}
\email{}

\author[0000-0001-6038-1923]{Kristopher G. Klein}
\affiliation{Lunar and Planetary Laboratory, University of Arizona, Tucson, AZ, USA}
\email{kgklein@arizona.edu}

\author[0000-0001-5258-6128]{Jasper S. Halekas}
\affil{Department of Physics and Astronomy, 
University of Iowa, 
Iowa City, IA 52242, USA}
\email{jasper-halekas@uiowa.edu}

\author[0000-0001-7379-4268]{Sujan Prasad Gautam}
\affiliation{Department of Physics and Astronomy, University of Delaware, Newark, DE 19711 USA}
\email{}

\author[0000-0001-7224-6024]{William H Matthaeus}
\affiliation{Department of Physics and Astronomy, University of Delaware, Newark, DE 19711 USA}
\email{}

\author[0000-0002-6962-0959]{Riddhi Bandyopadhyay}
\affiliation{Department of Physics and Astronomy, University of Delaware, Newark, DE 19711 USA}
\email{}

\author[0000-0003-0896-7972]{Srijan Bharati Das}
\affiliation{Center for Astrophysics $|$ Harvard $\&$ Smithsonian, Cambridge, MA, 02138, USA}
\email{}

\author[0000-0001-6692-9187]{Tatiana Niembro}
\affiliation{Smithsonian Astrophysical Observatory, Center for Astrophysics $|$ Harvard $\&$ Smithsonian, Cambridge, MA, 02138, USA}
\email{tniembro@cfa.harvard.edu}

\author[0000-0002-5982-4667]{Christopher J. Owen}
\affiliation{Mullard Space Science Laboratory, University College London, Holmbury St. Mary, Dorking, Surrey, RH5 6NT, UK}
\email{}

\author[0000-0003-0937-2655]{Rungployphan Kieokaew}
\affiliation{University of Toulouse, CNES, CNRS, IRAP, Toulouse, France}
\email{}

\author[0000-0003-0519-6498]{Ali Rahmati}
\affiliation{Space Sciences Laboratory, University of California, Berkeley, CA 94720-7450, USA}
\email{rahmati@berkeley.edu}

\author[0000-0002-4559-2199]{Orlando M. Romeo}
\affiliation{Space Sciences Laboratory, University of California, Berkeley, CA 94720-7450, USA}
\email{}

\author[0000-0002-5456-4771]{Federico Fraschetti}
\affiliation{Center for Astrophysics $|$ Harvard $\&$ Smithsonian, Cambridge, MA, 02138, USA}
\email{}

\author[0000-0002-2381-3106]{Marco Velli}
\affiliation{University of California, Los Angeles, Los Angeles, California, USA}
\email{mvelli@ucla.edu }

\author[0000-0002-6276-7771]{Lorenzo Matteini}
\affiliation{Department of Physics, Imperial College London, London, SW7 2BW, UK}
\email{l.matteini@imperial.ac.uk}

\author[0000-0002-5699-090X]{Kristoff W. Paulson}
\affiliation{Center for Astrophysics $|$ Harvard $\&$ Smithsonian, Cambridge, MA, 02138, USA}
\email{}

\author[0000-0002-2559-0831]{Lidiya Ahmed}
\affiliation{Department of Physics, Harvard University, Cambridge, MA, USA}
\email{}

\author[0000-0002-1989-3596]{Stuart D. Bale}
\affil{Physics Department, University of California, Berkeley, CA 94720-7300, USA}
\affil{Space Sciences Laboratory, University of California, Berkeley, CA 94720-7450, USA}
\affil{The Blackett Laboratory, Imperial College London, London, SW7 2AZ, UK}
\email{}

\author[orcid=0000-0002-7287-5098, gname=Phyllis,sname=Whittlesey]{Phyllis Whittlesey}
\affiliation{Space Sciences Laboratory, University of California, Berkeley, 7 Gauss Way, Berkeley, CA 94720, USA}
\email{phyllisw@ssl.berkeley.edu}

\author[orcid=0000-0002-1573-7457, gname=Marc,sname=Pulup]{Marc Pulupa}
\affiliation{Space Sciences Laboratory, University of California, Berkeley, 7 Gauss Way, Berkeley, CA 94720, USA}
\email{pulupa@berkeley.edu}

\author[orcid=0000-0003-2409-3742, gname=Nour,sname=Raouafi]{Nour E. Raouafi}
\affiliation{Johns Hopkins Applied Physics Laboratory, Laurel, MD 20723, USA}
\email{nour.raouafi@jhuapl.edu}

\author[0000-0002-7572-4690]{Timothy S. Horbury}
\affiliation{Department of Physics, Imperial College London, London, SW7 2BW, UK}
\email{}

\author[0000-0001-5127-9273]{Jean Morris}
\affiliation{Department of Physics, Imperial College London, London, SW7 2BW, UK}
\email{}

\author[0000-0002-9833-4097]{Helen O'Brien}
\affiliation{Department of Physics, Imperial College London, London, SW7 2BW, UK}
\email{}

\author[0000-0002-0396-0547]{Roberto Livi} 
\affil{Space Sciences Laboratory, University of California, Berkeley, CA 94720-7450, USA}
\email{}

\author[0000-0001-5030-6030]{Davin E. Larson}
\affil{Space Sciences Laboratory, University of California, Berkeley, CA 94720-7450, USA}
\email{}

\author[0000-0002-3020-9409]{Adam. J. Finley}
\affiliation{European Space Agency, ESTEC, Noordwijk, The Netherlands}
\email{adam.finley@esa.int} 

\author[gname=Philippe,sname=Louarn]{Philippe Louarn}
\affiliation{Institut de Recherche en Astrophysique et Plan\'etologie, CNRS, Universit\'e de Toulouse, CNES, Toulouse, 31028 CEDEX 4, France}
\email{philippe.louarn@irap.omp.eu}

\begin{abstract}
The early evolution of fast polar coronal hole (PCH) solar wind remains largely unconstrained by in situ measurements. In March 2025, Parker Solar Probe (Parker) at its closest approach of 9.86 Solar Radii ($R_\odot$) measured outflow from a large equatorial coronal hole (ECH) which was also measured at 1\,au and at intermediate distances by Solar Orbiter (also near its perihelion). At 1\,au the stream properties are consistent with PCH properties established by Ulysses. The stream was measured by Parker substantially below the Alfv\'en surface, with proton temperatures in excess of 2\,MK and a speed at $\sim$10\,$R_\odot$ which was only $\sim$60\% of its asymptotic value. The Solar Orbiter data indicates that the acceleration is largely complete by 60~$R_{\odot}$. Spherically-polarized fluctuations in the stream are observed to develop from near-transverse and small-angle at Parker to full reversal ``switchbacks'' at Solar Orbiter. Comparison of the implied acceleration profile to historical doppler-dimming measurements suggests that the stream's low coronal acceleration is similar to that of PCH flows. Consistent with previous work, this acceleration requires significantly more energy than can be provided by the observed thermal pressure gradients, with a significant contribution likely from the abundant Alfv\'enic fluctuation energy observed at Parker. These observations provide unique constraints on models of the radial evolution of the fastest solar wind, and indicate that these wind streams experience gradual, steady acceleration over their first few tens of solar radii of evolution. 
\end{abstract}


\section{Introduction} 

Fast solar wind streams are known to originate from coronal holes \citep{Nolte1976}, with an asymptotic speed at 1\,au that is positively correlated with source coronal hole size \citep[e.g.,][]{Nolte1976, Wang1990,Arge2003, Miralles2004,Rotter2012, Garton2018, Hofmeister2018}. The fastest such streams consistently reach a velocity in the range of 750--800\,km\,s$^{-1}$.

At solar minimum, the fastest streams are omnipresent at high latitudes due to the two large coronal holes co-located with the solar rotational poles \citep{McComas2000}. As characterized by Ulysses observations over the poles at a distance range of 1.34--5.4\,au, the fastest streams have very low variance in plasma properties: solar wind speeds in the range 750-800~km~s$^{-1}$, proton number densities of 2--3 (1\,au/R)$^2$\,cm$^{-3}$, proton temperatures of 0.26\,(1\,au/R)\,MK  \citep[][]{McComas2000}, and electron temperatures around 0.15 (1\,au/R)$^{0.91}$\,MK  \citep{Phillips1996}. They also have homogeneous in-situ compositional properties such as alpha abundances around $4\%$, elemental composition similar to the Sun's photosphere in contrast to the large variability seen from closed-field regions, and Oxygen charge state ratio around $\sim0.1-0.2$ corresponding to a coronal electron temperature of $\sim1$\,MK \citep{vonsteiger2000, vonSteiger2011, Zhao_2014}. 

These properties make this type of wind a compelling and information rich candidate to understand solar wind and coronal physics because this low variability implies lower complexity and variation in underlying physical processes compared to other types of solar wind \citep[][]{Sanchez-Diaz2016,Huang-2025,Alterman2025}, and (through the composition information) suggests minimal transport effects in terms of the coronal information that survives into the solar wind. This was partially why, in its original scientific formulation, Parker Solar Probe \citep[Parker; ][]{Fox2016,Raouafi2023} was proposed as a polar orbiter, like Ulysses, which would reach a high inclination orbit via a gravity assist with Jupiter and directly fly over polar coronal holes. However, this mission architecture would have come at a price of a multi-year orbital period and an extremely rapid velocity at perihelion leading to very small potential measurement time in the corona. Ultimately, the mission was instead implemented in a near-equatorial place orbit with a period as short as 88~days \citep[][]{Fox2016,Guo2021}.

Although much rarer near the solar equatorial plane, wind reaching the saturated range of 750-800~km~s$^{-1}$ is regularly measured at 1\,au, especially during solar maximum and the declining phase of the solar cycle when the largest equatorial coronal holes (ECH) occur. Unlike at high latitudes, by the time these streams are measured far from the Sun, they have been significantly distorted by the development of stream interaction regions \citep[SIRs;][]{Belcher1971,Pizzo1978} in which the leading edge of the stream is impeded by preceding slower streams, while the trailing edge is rarefied by the reduced dynamic pressure of the stream following it which is a challenge to interpreting their radial evolution \citep[][]{Hofmeister2026}. Despite this, the very fastest wind that forms the peaks of these largest SIRs still maintains remarkably similar and low variability properties including plasma density, oxygen charge state ratio, and First Ionization Potential (FIP) bias \citep[see e.g., Figure 1. in][]{Rivera2025}, in stark contrast to slower wind streams. Further, MHD modeling of idealized coronal holes suggests that when an `extended plateau' in velocity is observed in an SIR, it indicates a portion of the stream which has not yet been affected by the stream interface \citep{Hofmeister2022}. These facts motivate that it is still possible to gain useful physical insight into the early evolution of polar coronal hole wind through studying such saturated fast streams at low latitudes, which we here refer to as polar coronal hole-\textit{like} (PCH-L) solar wind. 

In particular, how close to the Sun the bulk of the acceleration of these streams occurs is a key constraint on models of (open field) coronal heating and solar wind evolution as it essentially traces where and how quickly energy is converted into bulk kinetic energy of the plasma, and is highly connected to the location of the stream's sonic point \citep[e.g.][]{Leer-Holzer-1980-sonicpoint,Wang1990}. Pre-existing observational constraints on this type of wind at and below 10\,$R_\odot$ all rely on remote sensing methods. In particular, the only existing constraints on solar wind speeds near 10\,$R_\odot$ in PCH flows come from radio scintillation \citep[][]{Grall1996,Breen1996, Moran1997} and suggest that the acceleration of PCH flow is rapid, and essentially complete within $10-15\,R_\odot$. However, these measurements are sensitive to velocity and density fluctuations along the line of sight and may convolve the bulk flow and Alfv\'en speed \citep[][]{Harmon2005}.  The near-Sun solar wind has been shown by Parker to be ubiquitously full of large amplitude Alfv\'enic fluctuations, ``termed magnetic switchbacks'' \citep[][]{Bale2019,Kasper2019} which suggest such fluctuations are large at these distance ranges. It is therefore highly informative to test these historical measurements with direct in situ detection of the plasma in which the bulk flow, fluctuation fields and Alfv\'en speed are all mutually distinguishable.    

In this Letter, we develop such constraints and comparisons through examining direct in-situ measurements of one such PCH-L wind stream measured by Parker at $\sim$10\,R$_\odot$ deep below the Alfv\'en surface (the location at which the bulk proton speed first exceeds the local Alfv\'en speed), by Solar Orbiter at $\sim$60\,R$_\odot$, and in a familiar SIR distorted state at the Earth-Sun L1 Lagrange point by the \emph{Wind} spacecraft. In Section \ref{sec:obs} we present the coronal and solar wind context of the stream and characterize the in situ observations. In Section \ref{sec:results} we present analysis on 1) the evolution of large-scale fluctuations in the stream and their relation to magnetic switchbacks and 2) the radial evolution of the stream's bulk properties in historical and statistical context. In Section \ref{sec:discussion} we develop an interpretation of these results and close in Section \ref{sec:conclusion} with major takeaways and outstanding directions for future studies of this and other streams.

\section{Observations} \label{sec:obs}

\begin{figure*}
    \centering
    \begin{subfigure}[b]{0.47\textwidth}
        \centering
        \includegraphics[width=0.98\linewidth, height=0.9\toprowht,
                         ]{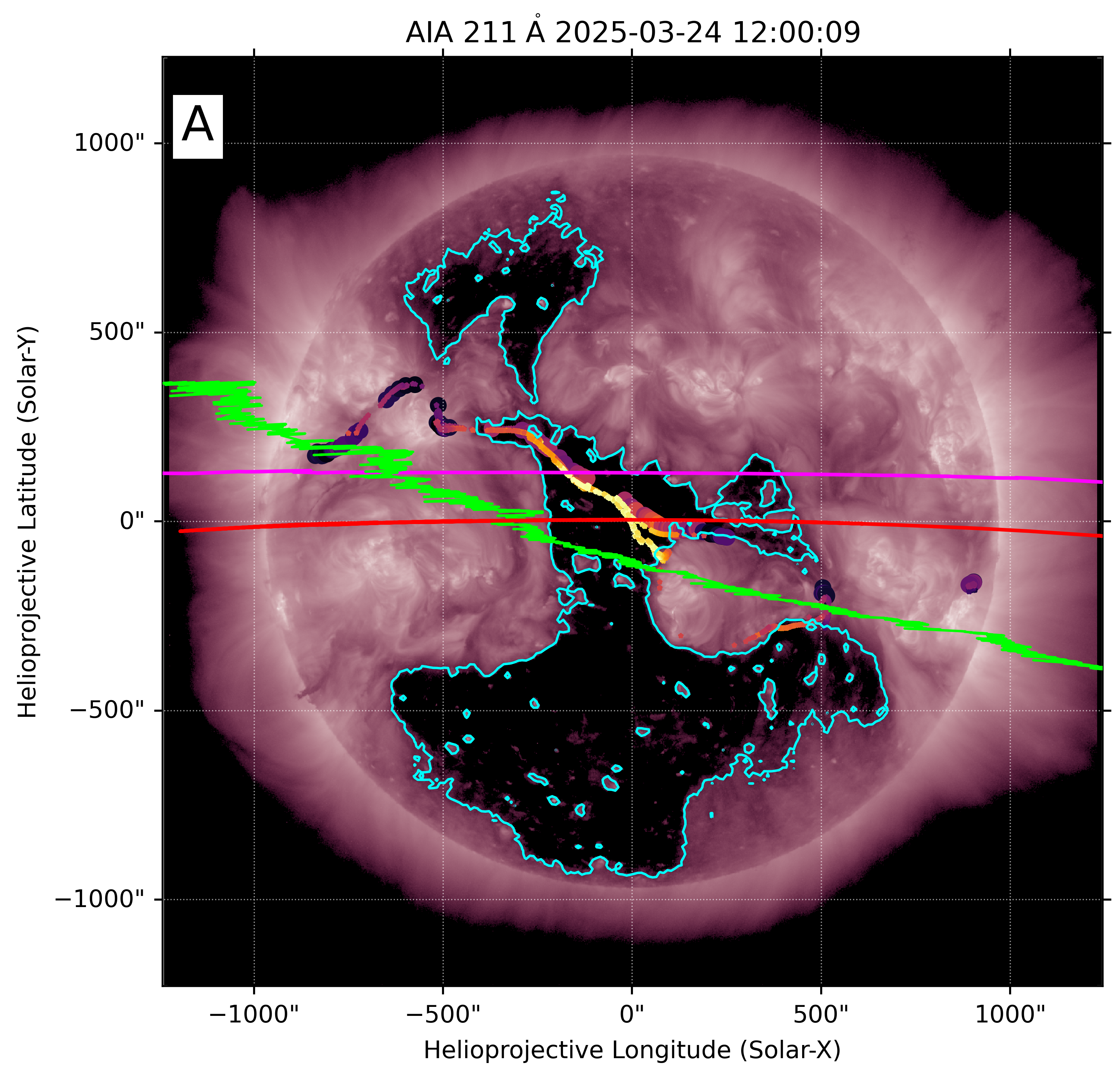}
        \label{subfig:A}\vfill
    \end{subfigure}
    \begin{subfigure}[b]{0.47\textwidth}
        \centering
        \includegraphics[width=\linewidth, height=\toprowht,
                         keepaspectratio]{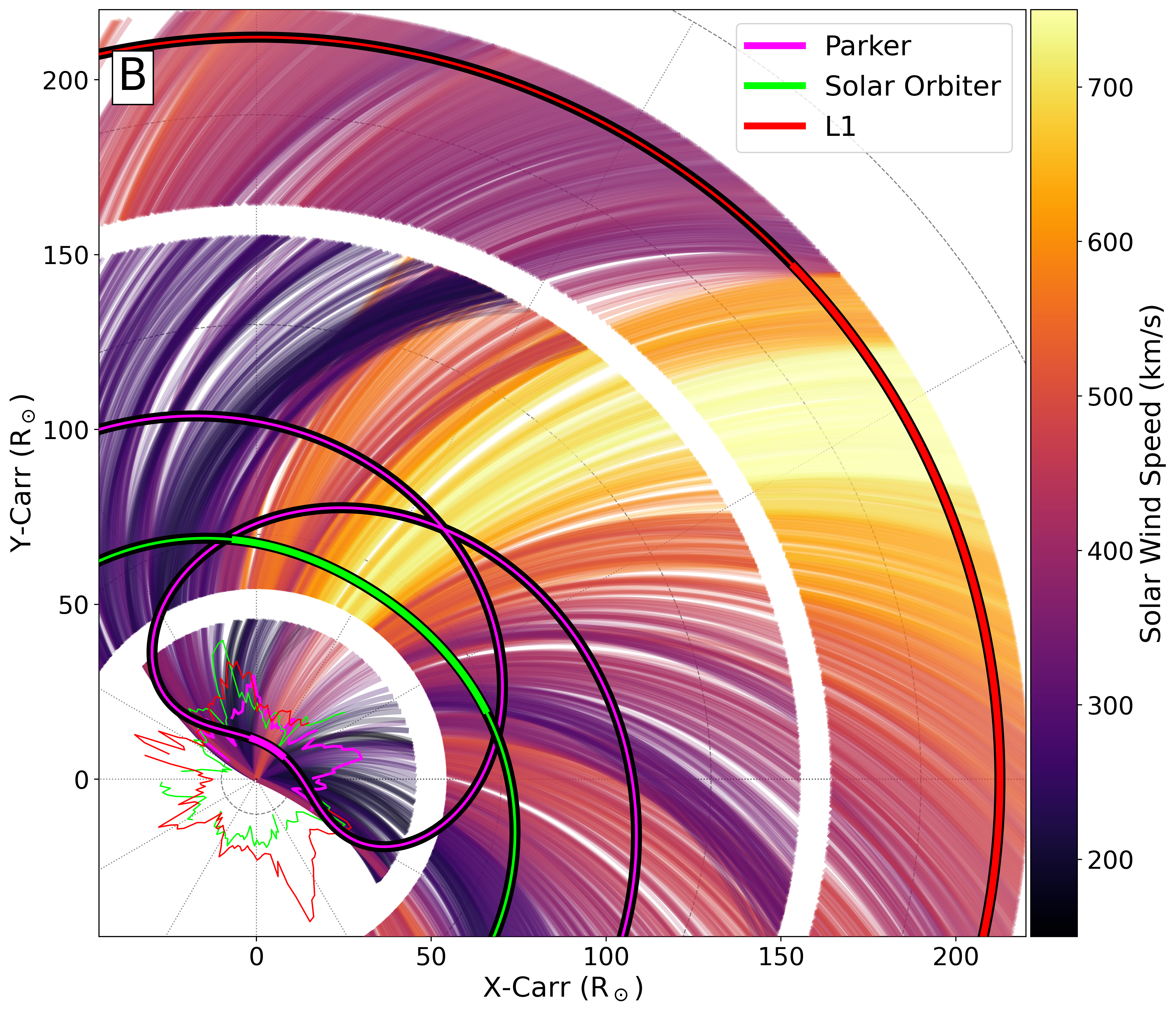}
        \label{subfig:B}
    \end{subfigure}\\%
    \begin{subfigure}[b]{1\textwidth}
        \centering
        \includegraphics[width=\linewidth, height=\botrowht,
                         keepaspectratio]{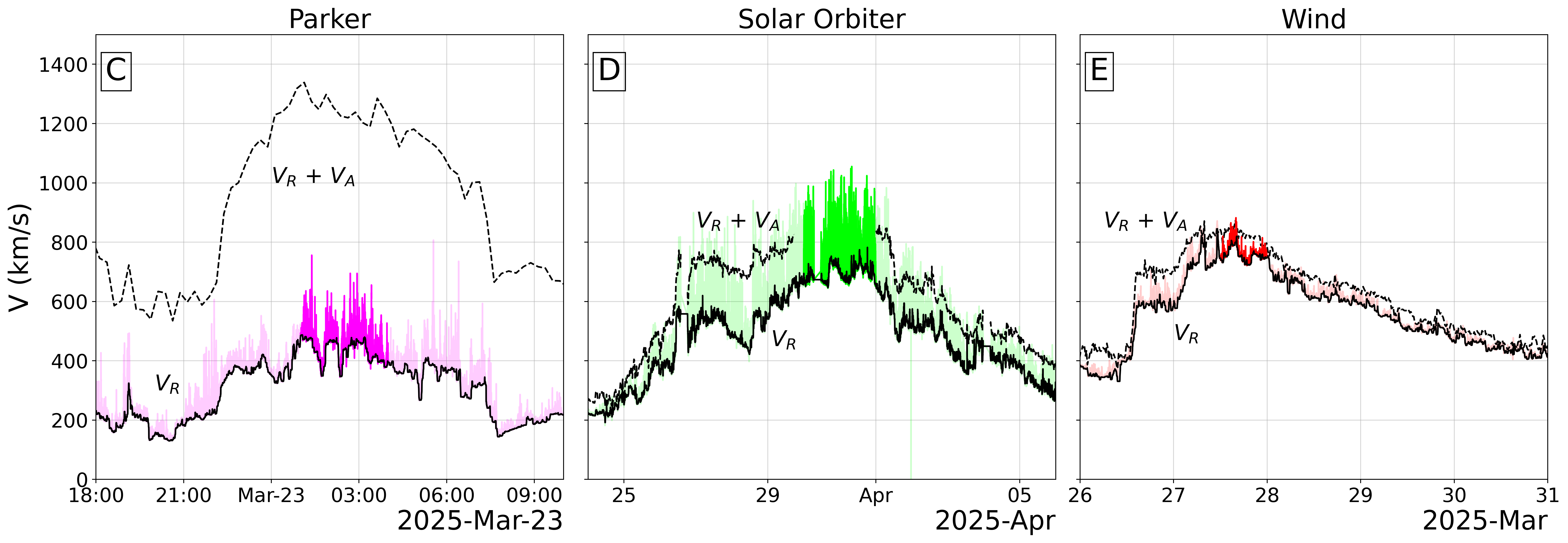}
        \label{subfig:C}
    \end{subfigure}%
    \caption{\textbf{Polar Coronal Hole-like fast wind stream measured at 3 points in the inner heliosphere.} 
    (A) An EUV 211~{\AA} image of the Sun on 2025/03/24 shows a large positive polarity coronal hole, outlined in cyan using a simple intensity threshold. The trajectories of Parker Solar Probe (magenta), Solar Orbiter (lime) and \emph{Wind} at the Earth-Sun L1 point (red) ballistically mapped to 2.5\,$R_\odot$ are projected onto the image. Resulting PFSS magnetic footpoints traced to 1\,$R_\odot$ are colorized by the measured velocity. The three sets of footpoints closely overlap and appear as a single trace. (B) Spacecraft trajectories plotted in the solar equatorial plane and used to draw Parker spiral lines using the measured velocity over annular ranges of distance corresponding to each spacecraft, and colorized by wind speed identically the footpoints in (A). Magenta/Lime/Red spiky curves close to the Sun show Alfv\'en surface shapes inferred from each spacecraft.
     (C-E) Radial proton velocity timeseries from each spacecraft. The native resolution measurements are shown colorized as with the trajectories in (A), median filtered baselines are plotted in black and the sum of the median velocity and Alfv\'en speed is overlayed in fainter black in each panel. The region corresponding to the peak of the stream is highlighted with a higher opacity.
     }
    \label{fig:fig1_stream_summary}
\end{figure*}

\subsection{Stream Source and In Situ Conjunction}\label{subsec:source_insitu}

Figure~\ref{fig:fig1_stream_summary} presents a full summary of the event studied in this work. Panel (A) shows that at the end of March, 2025, a huge equatorial coronal hole was observed on the Earth facing disk, as observed by SDO/AIA \citep[][]{Lemen2012}. Spacecraft trajectories ballistically mapped \citep[][]{Nolte1973} to 2.5\,$R_\odot$ are projected onto this Earth-centered view and overlaid in magenta (Parker,  projected from 10\,$R_\odot$), lime  (Solar Orbiter projected from 60\,$R_\odot$) and red (\emph{Wind}, projected from 1\,au). The spacecraft are all observed to cross the longitudes of the coronal hole at similar latitudes (despite the larger orbital inclination of Solar Orbiter). 
This magenta/lime/red color scheme will be followed throughout this work. Using these ballistically mapped locations, we trace coronal field lines from 2.5 to 1\,$R_\odot$ for all three crossings through a Potential Field Source Surface \citep[PFSS;][implemented through the sunkit-magex python package]{altschuler1969_pfss,Schatten1969_pfss,Stansby2020}, and driven by an ADAPT magnetogram \citep[][]{Arge2010,Hickmann2015} from March 24 2025. We scatter the resulting footpoints on the solar disk and colorize them according to the measured solar wind speed as the spacecraft cross these locations. All three footpoint traces are near identically located, and all three are observed to go through a local maximum in wind speed at the most interior point of the coronal hole that they sample. We therefore identify that fast wind from this large equatorial coronal hole are robustly sampled at three different radial locations in the inner heliosphere. 

Figure \ref{fig:fig1_stream_summary}B bolsters this case with additional context, presenting a complementary view projecting the measurement geometry in the solar equatorial plane in the Carrington frame. The respective spacecraft trajectories are plotted in the magenta/lime/red color scheme, and the specific spatial intervals corresponding to the fast stream are highlighted with thicker lines. Comparing the thick magenta, lime and red segments shows the relative distances and Parker spiral-shifted Carrington longitudes where the stream is measured. To further illustrate this, these measurement locations are used to seed Parker spiral streamlines generated and colorized according to the measured velocity identically to the footpoints in Fig.~\ref{fig:fig1_stream_summary}A. This illustrates that a Parker-spiral-aligned bright core corresponding the peak speed of the wind can be traced from Parker to \emph{Wind}; the ``brightening'' of this peak with distance is also clearly visible showing the significant acceleration from 10\,$R_\odot$ outwards which is the main subject of this work. The Alfv\'en surface geometry derived in \citet{Badman2025} \citep[see also][]{Finley2025} from these different spacecraft is also illustrated and lie in the radial range of the Parker measurements. In particular, Parker crosses this stream a significant distance below the Alfv\'en surface, while the Solar Orbiter and \emph{Wind} measurements are well above it.
Lastly for Figure~\ref{fig:fig1_stream_summary}B, this ballistic illustration  shows clearly how the slow (dark colors) wind preceding and trailing the stream respectively asymmetrically distort the stream leading to the leading compression, and trailing rarefaction consistent with stream interaction region (SIR) formation.

Figures~\ref{fig:fig1_stream_summary}C-E show the full resolution measurements of the radial proton speed taken by each spacecraft. Brighter colors indicate the region of the stream we identify as the pristine ``peak'' of the stream (see the following section~\ref{subsec:streamprops} for a discussion of our rationale in each case). A solid black line in each panel shows a median filter tracing the bottom of each speed profile, characteristic of intervals with outwards propagating spherically polarized Alfv\'en waves which manifest as 1-sided velocity spikes \citep{Gosling2009, Matteini2014, Kasper2019}. A dashed black line in each panel shows the sum of the (median) proton speed and Alfv\'en speed (with a gap in Panel D due to a lack of Solar Orbiter MAG data at this time). This construction will primarily be useful for the discussion developed later in $\S$~\ref{subsec:flucs}, but in brief, this line represents the radial velocity threshold corresponding to a greater than 90~degree rotations in the magnetic field vector \citep[i.e., $V_R$ crossing this line corresponds to  ``full-reversal'' switchbacks;][]{Badman2026} which are often used as a threshold for switchback identification \citep[e.g.][]{Macneil2020, Ruffolo2020, Pecora2022}. Cross-comparing these three panels, the significant acceleration of the stream is apparent, as is the overall deformation of the stream from nearly symmetric at Parker to the clear 1-sided pattern of SIRs that has developed by 1\,au. 

Note that due to the orbital dynamics, Parker samples the stream's spatial structure in the reverse time order to Solar Orbiter and \emph{Wind}, therefore the right side of the Parker timeseries is what evolves into an SIR compression and the left side is what evolves into a rarefaction. This will be more intuitive to compare in the next section where the streams properties are plotted as a function of Carrington longitude, rather than time.

\subsection{Average Stream Properties}\label{subsec:streamprops}

\begin{figure*}
    \centering
    \includegraphics[width=\linewidth]{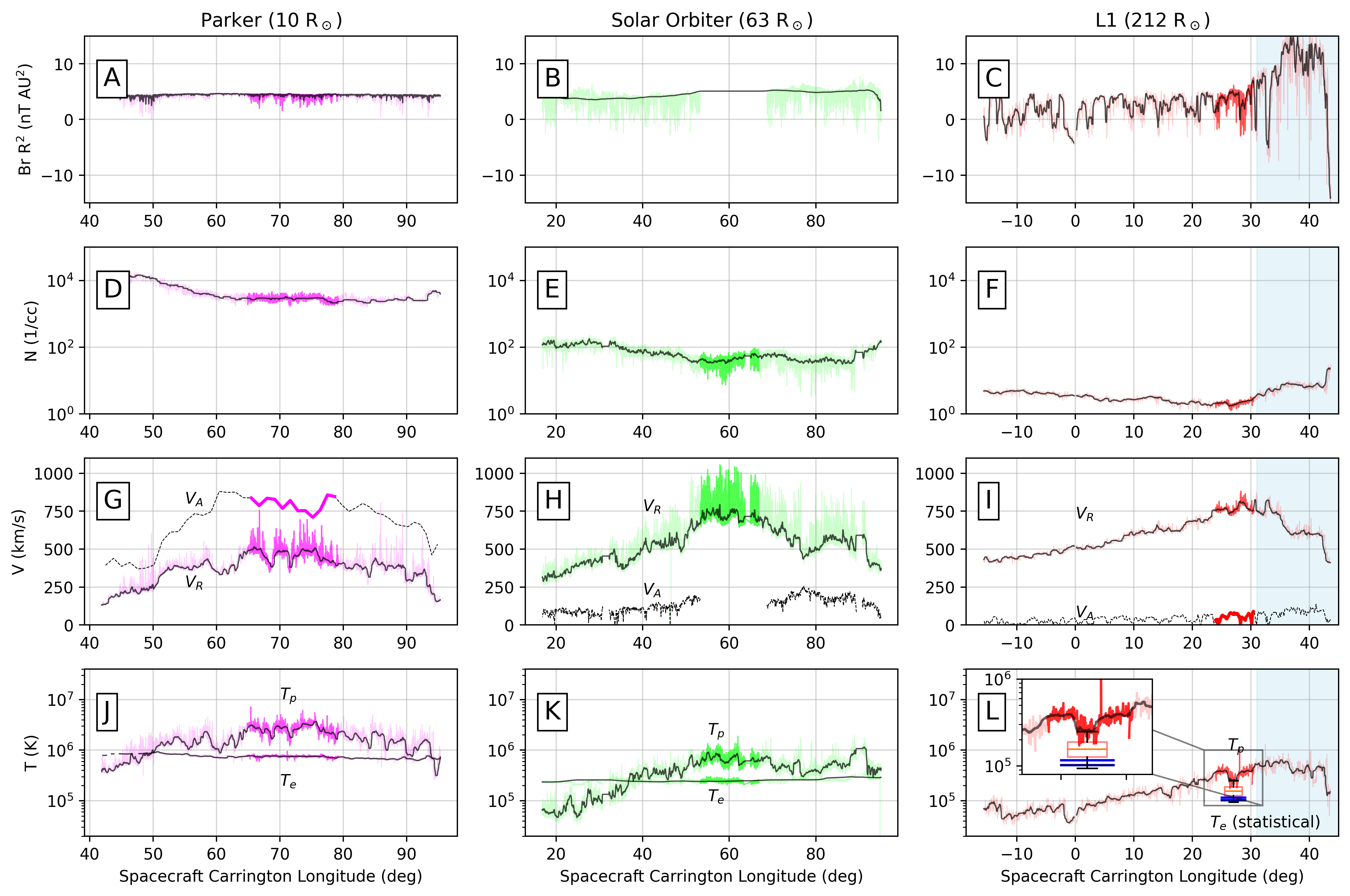}
    \caption{\textbf{In situ measurements of the bulk stream properties as a function of heliographic longitude.} Left hand column: Parker measurements around 10 R$_\odot$ showing (from top to bottom) the radial component of the magnetic field, the plasma density, the bulk proton speed and Alfv\'en speed, and the proton and electron temperatures. All data are direct measurements of the stream except for $T_e$ at 1\,au (bottom left) where various statistics for fast wind from \emph{Wind} \citep[][box plot and light blue]{Wilson2026} and Ulysses \citep[][dark blue]{Phillips1996} are presented. An inset is provided to better differentiate these data. As in figure \ref{fig:fig1_stream_summary}(C - E), the full resolution data is shown in color with a median filtered baseline in black, and a sub-range of longitude capturing the peak of the stream is highlighted with higher opacity. The middle and right hand columns show the corresponding properties measured at 60\,R$_\odot$ by Solar Orbiter and near 1\,au by \emph{Wind}. In the right hand column, faint blue shading shows a region identified as SIR-dominated (see main text).}
    \label{fig:fig2_stream_properties}
\end{figure*}

Figure~\ref{fig:fig2_stream_properties} presents the overall stream properties at each measurement location as a function of spacecraft Carrington longitude.\footnote{i.e., the Carrington coordinate of the spacecraft location, not its projected location at 2.5\,$R_\odot$} Each row of panels shows the same quantity measured with increasing heliocentric distance from left to right. As in Fig.~\ref{fig:fig1_stream_summary}C-E, the peak of the stream is highlighted in brighter colors, and a black line shows a median ``background'' (binned across a few tenths of a degree in spacecraft Carrington longitude) for each quantity.

In addition to the solar wind speed presented previously, these panels show the solar wind magnetic field strength (normalized to nT at 1\,au), the plasma density, the Alfv\'en speed, and the proton and electron (where available) temperatures. A full accounting of the different instruments from which these data are derived along and associated caveats and limitations are presented in Appendix \ref{appendix:instruments}. Due to spacecraft operations, no Solar Orbiter magnetic field data was available at the peak of the stream so a data gap is apparent in $B_R R^2$ and the Alfv\'en speed in the respective Solar Orbiter panels. 

From the top row, the radial component of the magnetic field normalized by $1/R^2$ shows each measurement corresponds to positive magnetic polarity solar wind, and magnetic flux is largely flat and conserved (i.e. scales as $1/R^2$). The exception to this is at 1\,au (Panel C), where the compression of the SIR at 1\,au starts to become significant. 

The second row shows that the plasma density decreases (approximately as $1/R^2$, but more accurately according to conservation of mass flux, as will be shown in Sec. \ref{sec:results}) across distance and is relatively flat, although weakly identifiable as a local minimum. At 1\,au, as with the magnetic field strength, the density increases in the SIR compression region.  

As discussed previously, the solar wind speed strongly accelerates from $10\,R_\odot$ where the peak of the baseline speed is 450--500\,km\,s$^{-1}$, to $63\,R_\odot$ where it touches 750\,km\,s$^{-1}$ and beyond to 1\,au where it is in the range of 750--800\,km\,s$^{-1}$. The Alfv\'en speed correspondingly decreases from a staggering 800\,km\,s$^{-1}$ at Parker (far exceeding the solar wind speed, with Alfv\'en Mach number $M_A\sim0.6$), to under 50\,km\,s$^{-1}$ at 1\,au. Comparing the median velocity trace to the instantaneous measurements, we observe that the peak of the stream has abundant one sided velocity spikes \citep[][]{Kasper2019} at both Parker and SolO. These regions of velocity spikes are separated by short and rapid decreases in speed, which, in this stream, exhibit reduced magnetic footpoint field variability and lower magnetic expansion factors in comparison to the switchback patches \citep[][]{Jaffarove2026}. 

Interestingly, the spikes appear of similar amplitude at SolO and Parker while almost negligible at \emph{Wind}, meaning that the maximum instantaneous amplitude of V$_R$ actually peaks around Solar Orbiter distances and not closer in. However, as will be shown in section \ref{subsec:flucs}, this should not be confused with the overall amplitude of the fluctuations remaining constant because what is plotted here is only a 1D (radial) projection of the fluctuations and the non-radial fluctuation components at Parker are much larger than at SolO.

Lastly, in the bottom row we examine the proton and electron temperatures. The proton temperature largely tracks the velocity structure of the stream in each case and at the peak of the stream is higher than the associated electron temperature, which remains flatter throughout. The peak proton temperature at 10\,$R_\odot$ is around 3\,MK (i.e. in the coronal temperature regime), while the electron temperature is around 0.8\,MK. Both species' temperatures decrease steadily with distance from the Sun. Due to uncertainty in the spacecraft potential at \emph{Wind}, we use statistical proxies for the electron temperature based on prior work from \emph{Wind} data at L1 \citep[][box plot, and lighter blue line - see also Appendix \ref{appendix:Te}]{Wilson2023,Wilson2026} and from Ulysses data from PCH wind \citep[][darker blue line]{Phillips1996}.

Figure~\ref{fig:fig2_stream_properties} also illustrates our rationale for choosing the ``peak'' of this stream, which may be summed up as a maximum plateau in wind speed and proton temperature, and a coincident minimum plateau in density. This is relatively straightforward and insensitive at Parker and SolO, but at \emph{Wind}, we note that if only judging by a plateau in velocity, the region would include a sharp increase in magnetic flux, density and proton temperature. These are all indicators of the development of an SIR compression region from the exchange and interaction of the fast wind with neighboring slow. We therefore use these non-velocity measurements to shorten the region of the velocity plateau which we identify as the pristine peak of the stream. The excluded compression region is shaded in light blue in the right hand column of Figure~\ref{fig:fig2_stream_properties}. 

We have therefore presented the full context and rationale for the identification of this stream as ``polar-coronal hole like" (PCH-L), as well as our best effort at isolating a pristine peak whose properties can be confidently studied with distance. 

\section{Results}\label{sec:results}

Here, we present the main results we obtain by studying the properties of this PCH-L stream at 10\,$R_\odot$ and how it evolves out to 1\,au. We first (Sec. \ref{subsec:flucs}) revisit the Alfv\'enic fluctuations present in the stream and comment on how they relate to ``magnetic switchbacks'' \citep[][]{Bale2019,Kasper2019}. We then (Sec. \ref{subsec:RadialEv}) place the radial evolution of the bulk properties in statistical and historical context, and discuss the implications for the early evolution of fast wind flows in general.  

\subsection{Fluctuations and Switchback Evolution}\label{subsec:flucs}

\begin{figure*}
    \centering
\includegraphics[width=0.7\linewidth]{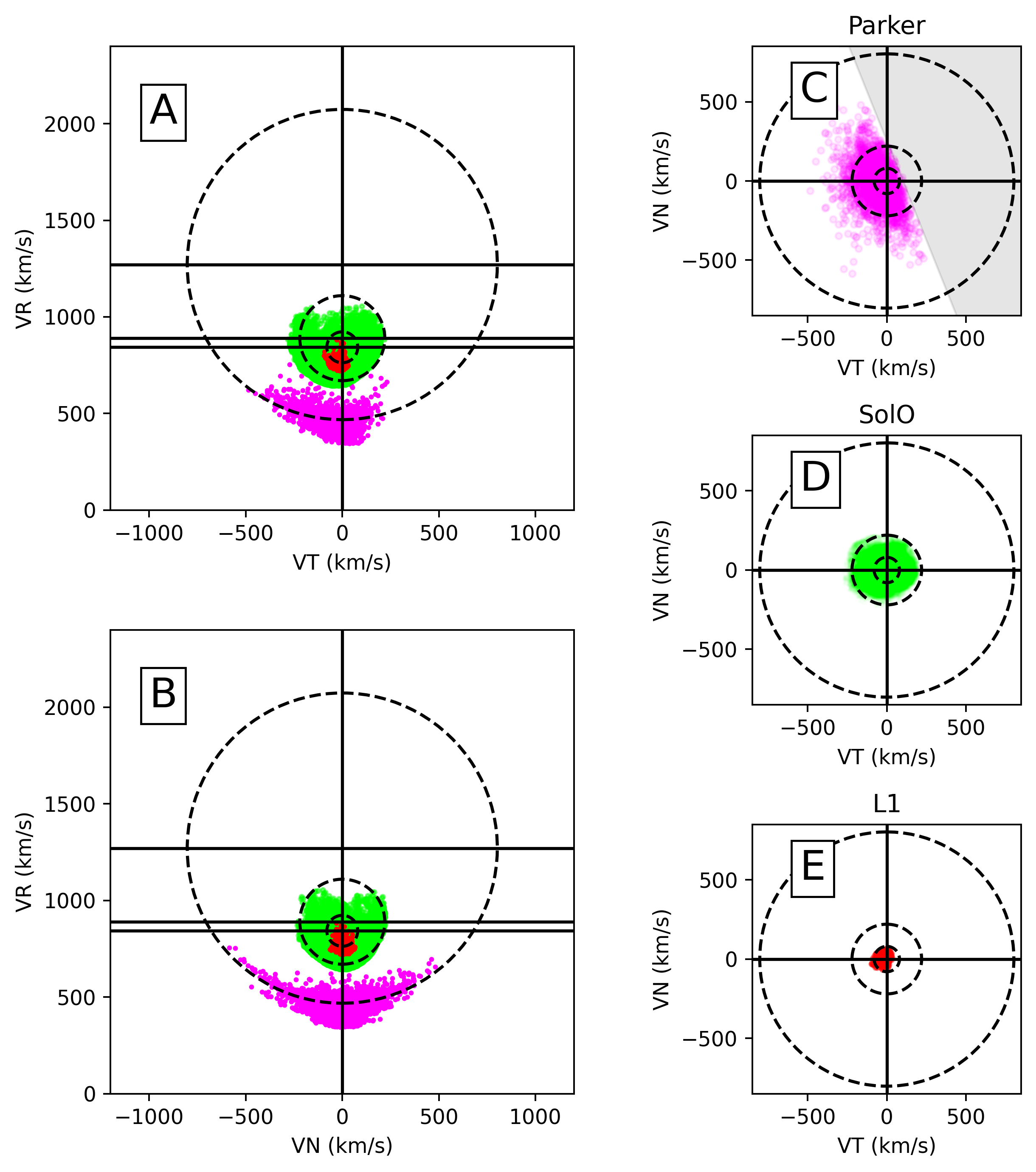}
    \caption{\textbf{Evolution of velocity fluctuations in the fast wind stream}. Measured vector components of the proton velocity measured by Parker (magenta), Solar Orbiter (lime) and \emph{Wind} (red) are displayed in various combinations. Left column: $V_R$ vs $V_T$ (top) and $V_N$ (bottom) are displayed for all three spacecraft measurements, with their respective `Alfv\'en spheres' overlaid as dashed circles. Solid cross hairs indicate the centroid of each sphere, which is interpreted to be the stream's De Hoffman-Teller velocity. Right column: Corresponding plots of $V_N$ vs $V_T$, showing the projection of the fluctuations perpendicular to the radial direction. These are separated by spacecraft from top to bottom, avoiding overlap of the data points, with all three Alfv\'en sphere projections displayed in each case.In the top right panel, grey shading indicates the region of velocity space outside the SPAN-i field of view at this time and is observed to accurately correspond to the apparent diagonal cutoff in the Parker V$_N$-V$_T$ data (magenta).}
    \label{fig:fig3_switchbacks}
\end{figure*}

Figure \ref{fig:fig3_switchbacks} shows the 3D evolution of the velocity fluctuations from the peak of the stream. The third row of Figure \ref{fig:fig2_stream_properties} showed only the radial projection of these fluctuations, but here we show all 3 vector components. The two panels in the left column show $V_R$ vs $V_T$ (the direction of the Sun's rotation) and $V_R$ vs $V_N$ (the out-of-plane component) and superimpose the measurements from Parker, Solar Orbiter and \emph{Wind}. Additionally we annotate a dashed circle for each spacecraft with radius given by the median Alfv\'en speed of the stream\footnote{Except for Solar Orbiter for which magnetic field data is not available at this time. Here instead, we size the radius of the sphere based on the spherical locus the fluctuations draw in velocity space which can be understood as an indirect ``field-free'' estimate of the Alfv\'en speed.}, and a center given by the median of $V_R + V_A$, which approximately corresponds to the zero motional electric field (or de Hoffmann-Teller) frame. We term this construction the `Alfv\'en sphere' which is observed to be a good descriptor of the spherical polarization \citep[][]{Matteini2015,McManus2022,Bowen2025, Ding26, Sioulas26} of the fluctuations of the velocity vector. 

Equivalent panels in the right-hand column show the $V_T-V_N$ components. Since in this projection the measurements' Alfv\'en spheres and measurement populations strongly overlap, we separate the spacecraft measurements into different panels but keep all three Alfv\'en sphere annotations for reference.

Comparing across spacecraft, we observe the Alfv\'en sphere shrink (with the Alfv\'en speed), the lower extent of the sphere translate upwards (corresponding to the acceleration of the bulk speed), and the center of the sphere, which is approximately equivalent to the de Hoffmann-Teller frame, decrease and approach the bulk speed. The portion of the Alfv\'en sphere which is populated also evolves strongly with distance. For Parker at 10\,$R_\odot$, we see that although the radial velocity spikes shown in Figure~\ref{fig:fig2_stream_properties} were not particularly impressive, this actually corresponded to a radial projection of small angle fluctuations of the 3D vector whose tangential and normal components were dominant. This means that although the spherical polarization is clearly exhibited, the fluctuations are closer to transverse in nature (which would be the limit for infinite Alfv\'en speed). 

At Solar Orbiter, we observe this picture has changed dramatically with nearly the entire sphere (which is now much smaller in velocity space) being populated. Although not shown here, these velocity fluctuations are strongly Alfv\'enically correlated to the magnetic field fluctuations \citep[$\sigma_c \sim 1$ at Parker, see][]{ Bandyopadhyay2026_inrev}. This means that when the velocity fluctuations rotate more than 90$^\circ$ on their Alfv\'en sphere, the corresponding magnetic field fluctuations invert in polarity and become ``full reversal switchbacks'' (deflection of the magnetic field $>90^\circ$). This connection between large B and V vector evolution in the case of spherical polarization are well-described by \citet[][see their Fig 4 and Eqn 2]{Matteini2014}.

We therefore see direct evidence that this ``switched back'' state develops through radial evolution between Parker and Solar Orbiter from fluctuations which were originally smaller angle but still spherically polarized. We also note that although the radial projection (Fig.~\ref{fig:fig2_stream_properties}) of the fluctuations makes them appear to stay relatively constant in amplitude from Parker to Solar Orbiter \citep[][]{Ding26}, the full vector picture confirms that the fluctuations have still decreased significantly in amplitude and lost energy, but appear more dramatic at Solar Orbiter because $|d \textbf{V}|$/$V_A$ has increased. This overall increase in the spectrum of switchback deflection angles with distance or Alfv\'en Mach number has been shown statistically from the magnetic field data in recent work \citep[][]{Pecora2022, Bowen2025, Goodwill2026, Payne2026, Mallet2026}. 

This evolution continues from $60\,R_\odot$ out to 1\,au, where the Alfv\'{e}n sphere continues to shrink and remains fully populated. We see an additional effect where the $T$ and $N$ components behave differently, with the distribution of velocity measurements at Solar Orbiter, and more so at \emph{Wind}, becoming skewed in the $-T$ direction while remaining evenly distributed in $N$. We interpret this as a likely 2D effect from the development of the stream interaction rarefaction inducing a $-T$-directed thermal pressure gradient on the stream of the peak, and has previously been identified as a complicated in directly measuring angular momentum flux in the inner heliosphere \citep[][]{FInley2019}. At first observation, it may also appear that the same $T$-$N$ asymmetry is present in the Parker measurements: the distribution of $V_N$ is evenly distributed about zero but in $T$ it is sharply cut off in the $+T$ direction. However, in this case, this is attributed to measurement effects: The SPAN-i sensor on Parker has a field of view (FOV) cutoff due to the spacecraft heat shield impeding the apparent velocity directions which can reach the detector \citep[][]{Livi2022} and limiting the maximum tangential flow which can be detected \citep[][]{Badman2023_AGU}. For the current interval, even though the relative speed of the spacecraft at perihelion is large enough that radial-directed flows are comfortably detected, the enormous tangential velocity deflections in these spherically polarized Alfv\'en waves is sufficient to move the protons out of the detectable range when the deflection is primarily in the $+T$ direction. The actual cutoff is slightly tilted in radial, tangential and normal ($RTN$) coordinates due to the relative orientation of the instrument, but is clear to see when observing $V_N$ vs $V_T$ in the top right panel. The shaded region shows the ranges of velocity vector components which are not accessible to SPAN-i for the known spacecraft aberration velocity at this time. It forms a diagonal boundary which very accurately describes an apparent ``edge" in the measurement population, which can therefore be identified as fully artificial. 

\subsection{Radial Evolution of Stream Bulk Properties}\label{subsec:RadialEv}

\begin{figure*}
    \centering
    \includegraphics[width=0.9\linewidth]{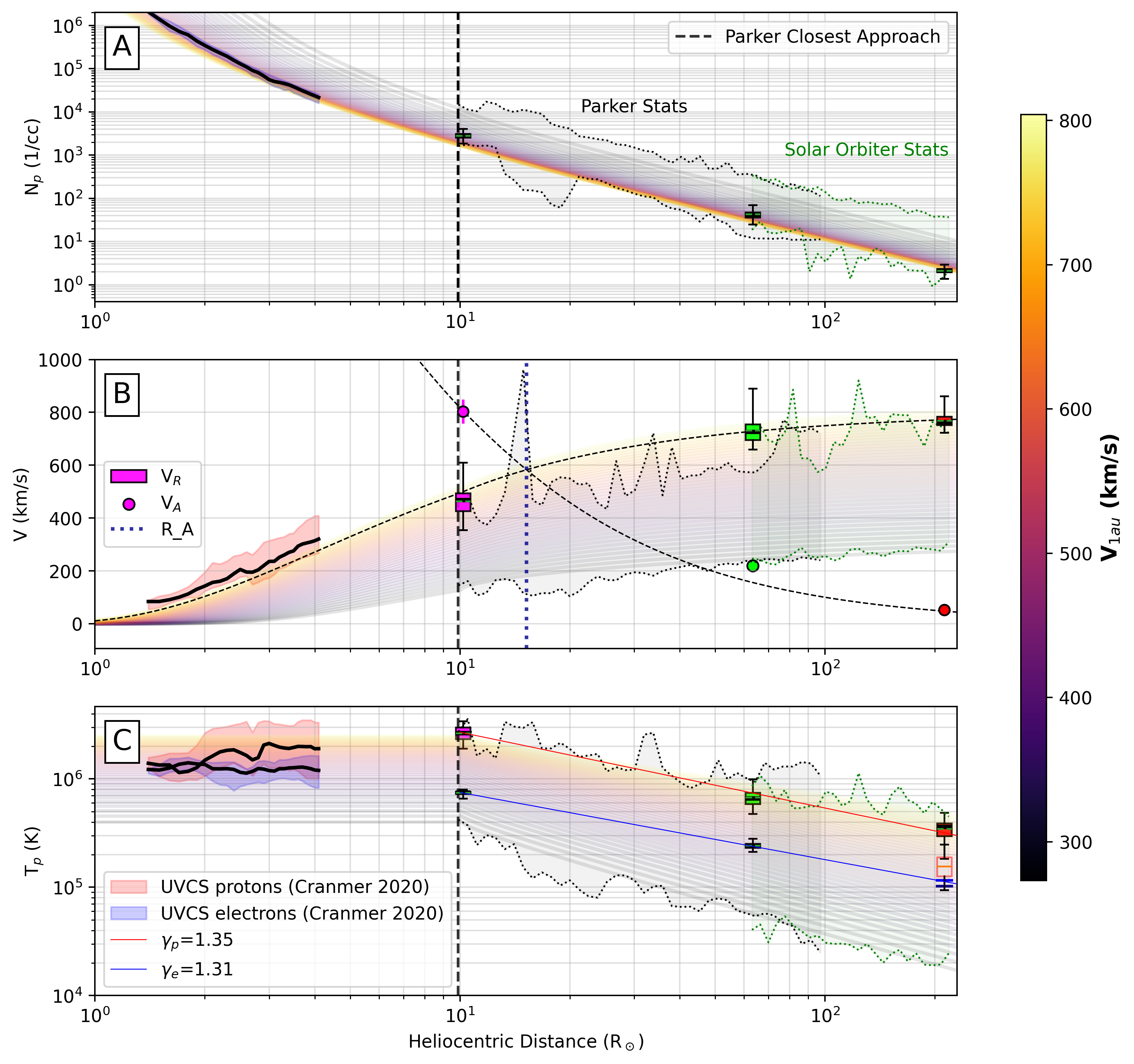}
    \caption{\textbf{Properties of the peak of the stream as a function of heliocentric distance in historical and statistical context.} Panels A-C show plasma density, proton and Alfv\'en speed (middle), and proton and electron temperatures. Each panel shows a boxplot of the Parker (magenta), Solar Orbiter (lime) and \emph{Wind} measurement (red) of the respective property, where the boxplot whiskers are chosen to span the 1st and 99th percentiles of the peak measurements. In the background, a statistical data set  of measurements from Parker (black/grey) and Solar Orbiter (greens) are draw the 1st and 99th percentiles in different distance bins (faint shading, bounded by dotted curves). Additionally, a set of “iso-poly” curves with external forcing \citep[][]{Dakeyo2022,shi2022_polyt,Rivera2024,Rivera2025}  which spans the dataset \citep[][]{Badman2025} is overlaid and colorized from yellow to purple according to the speed at 1\,au and the best matching velocity and related Alfv\'en speed profile highlighted in (B) with black dashed curves. A vertical blue dashed line in B shows the estimated Alf\'en critical point for this stream. Lastly, black lines and red/blue (proton/electron) shading indicate the median and interquartile ranges of corresponding PCH properties inferred from UVCS at low coronal heights \citep{Cranmer2020a}.}
    \label{fig:fig4_radialevolution}
\end{figure*}

Lastly we present and contextualize the radial evolution of the PCH-L stream's bulk properties as directly measured at $10\,R_\odot, 60\,R_\odot$ and 1\,au.

In Figure \ref{fig:fig4_radialevolution}, we show the statistics of the measurements identified as the peak of the stream in Figure \ref{fig:fig2_stream_properties} as box plots showing the measurement median and interquartile range (box) and the 1st-99th percentiles (whiskers), plotted against heliocentric distance in log-space. We also overlay a family of ``iso-poly + forcing'' curves \citep[][blue to yellow curves, colored according to the solar wind speed at 1\,au]{Dakeyo2022,Shi2022,Rivera2024,Rivera2025} with parameters sized to span the 1st-99th percentile of measurements as a function of heliocentric distance provided by Parker and Solar Orbiter to date \citep[dotted black curves, see][]{Badman2025}\footnote{We note an apparent enhancement in the 99th percentile of wind speed at 13-14 $R_\odot$. We attribute this to anomalously fast wind observed during a large CME in Encounter 13 \citep[][]{Romeo2023}. This distance range should be monitored as statistics are added in the future, to verify this exception continues to smooth out.}. A representative fast wind profile is highlighted (black dashed curves) in panel (B). To compare these model curves' low-coronal behavior, we plot from 1.4--4\,$R_\odot$ measurements of PCH flows at solar minimum as measured by the Ultraviolet Coronagraph Spectrometer \citep[UVCS;][]{Kohl1995} as reported by \citet[][]{Cranmer2020a}. The panels show plasma density (Panel A), bulk (proton) and Alfv\'en velocity (Panel B) and proton and electron temperatures.

We observe a mass-flux conserving decrease in density and significant acceleration from 10 to 60\,$R_\odot$ with a small remaining residual acceleration from 60\,$R_\odot$ to 1\,au. This corresponds to cooling in protons and electrons which is well described by a single power law through all three measurement points with polytropic indices $\gamma_p=1.35, \gamma_e=1.31$. We observe all these properties lie very close, at all distances, to the fastest (by 1\,au speeds) iso-poly model (dashed black curves, panel B), fitting the qualitative historical picture \citep[][]{McComas2000} that PCH (and here PCH-L) wind is uniformly rarefied, fast and has the highest proton temperatures across all wind types. We also observe that these measurements lie at the 99th (1st) percentile of all temperature and velocity (density) measurements from Parker and SolO at each distance, and is therefore a very rare measurement in the current statistical context of in situ data of the most recent solar cycle in the inner heliosphere. 

Comparing the Parker measurements to the crude isothermal behavior of the isopoly models and the UVCS constraints below $10\,R_\odot$, the picture remains relatively well behaved. The extrapolated acceleration profile, and associated departure of the density from $1/R^2$ behavior matches the proton acceleration, Alfv\'en speed and density constraints. The Alfv\'en speed for Solar Orbiter here is an indirect estimate from the spherical polarization due to the lack of magnetic field data, and is seen to slightly overestimate the expected trend which matches the Parker to L1 evolution very well. 

The proton temperature (Panel C) at Parker is slightly in excess of the UVCS proton temperature lower down, which is consistent with the UVCS suggestion of a weakly increasing proton temperature in the corona. This suggests this Parker measurement is at or close to the proton temperature maximum for this stream, although synchronous determinations of proton temperatures \citep[which are not currently available;][]{Rivera2025_IAU} would be required to fully constrain this. This also suggests a weakly increasing proton temperature in the corona may be a better approximation to this stream than the pure isothermal assumption made here. The electron temperature at Parker ($\sim$0.8~MK) is slightly below the 1~MK constraints from UVCS. The UVCS electron temperature constraint indicates isothermal behavior is a good approximation out to somewhere in between 4\,$R_\odot$ and $\sim$10\,$R_\odot$, so a resolution to this may be that the polytropic cooling of the electrons extends closer in than Parker while the proton temperature turns over. Both these generalizations beyond an isothermal corona have been implemented recently \citep[][]{Dakeyo2025_bipoly} and comparisons and parameter fitting with this dataset would be a compelling future investigation. 



\section{Discussion}\label{sec:discussion}

Parker Solar Probe perihelion 23 in late March 2025 cut through a nascent polar coronal hole-like (PCH-L) stream around 10\,$R_\odot$. The same stream is readily identified in several crossings further out which show that its asymptotic speed is in the range of 750--800\,km\,s$^{-1}$ at 1\,au. In Figure~\ref{fig:fig1_stream_summary}A we showed that the source was unambiguously traceable to one of the largest equatorial coronal holes of solar cycle 25 to date. It is also aligned with the solar magnetic dipole axis, which was dominating the coronal structure at this time \citep[][]{Rivera2025_AGU} adding to its similarity with the solar rotation axis-aligned polar coronal holes at solar minimum.

A survey of the peak properties at 1\,au, yields a speed of $763_{-11}^{+11}$\,km\,s$^{-1}$, a plasma density of $2.1_{-0.2}^{+0.2}$\,cm$^{-3}$ and a proton temperature of $0.36_{-0.6}^{+0.3}$\,MK  (where each quoted measurement here and below conveys the median and interquartile range). These are all squarely in the range of values observed by Ulysses over the Sun's rotational poles at solar wind \citep[e.g.][]{Phillips1996, McComas2000}, so we establish that this stream has all the asymptotic properties to term it PCH-L.  

While limited observations of such streams reaching the same asymptotic properties have been made in the lower and middle corona with remote sensing techniques, including UVCS, and radio scintillation, the in situ observations presented here are more comprehensive and unambiguous, avoiding issues such as line of sight integration. These data provide multiple direct diagnostics and key constraints on the early evolution of flows from large coronal holes. Critical for future studies, these measurements include concurrent magnetic field and fluctuation data which is extremely challenging to obtain from remote observations at comparable heights.

The multi-spacecraft measurements at different distances show that while the boundaries of the stream evolve strongly, causing a typical SIR signature by the time it reaches 1\,au, the middle of the stream is relatively non-interacting and therefore its radial evolution can be probed \citep[see][for further examination of evolution of different parts of the SIR]{Payne2026_inprep}. We buttress this claim by not only relying on the identification of a plateau in velocity \citep[][]{Hofmeister2022}, but also rejecting portions of the SIR where magnetic compression and heating are detected in the stream (Fig \ref{fig:fig2_stream_properties}). 

Further, by observing that the streams velocity fluctuations populated a sphere of near constant radius corresponding to the local Alfv\'{e}n speed (discussed further below), a well established bulk wind speed could be separated from the one-sided fluctuations \citep[Fig.~\ref{fig:fig3_switchbacks}, see also][]{Gosling2009, Ding26} leading to an accurate measurement of the stream's  acceleration profile. For spherically polarized fluctuations, we emphasize that this baseline speed is better isolated through median averaging since the mean results in an average velocity that is not a member of the measurement ensemble \citep[it lies in the interior of the sphere, see also][]{Badman2021,Bandyopadhyay2026_inrev}.  

We characterize this PCH-L stream at 10\,$R_\odot$ and determine that it is deeply sub-Alfv\'enic ($M_A\sim$0.6), is only at $60\%$ of the speed it attains by 1\,au ($468_{-41}^{+26}$\,km\,s$^{-1}$) and is measuring ion temperatures which are in excess of 2\,MK ($2.58_{-0.40}^{+0.38}$\,MK), and likely near the ion temperature maximum for the stream. The electron temperature is lower ($\sim$0.8\,MK) but also approaching typical coronal hole source region temperatures \citep[][]{Doyle2010, Nistico2011, Cranmer2020b, Hegde2024}. We therefore suggest this measurement constitutes a sample of a fast wind stream in a stage of its evolution which may fairly be called ``coronal'' in nature. The stream is also extremely rarefied ($2900_{-300}^{+200}$\,1/cc), near the 1st percentile of measurements at this distance, and has an Alfv\'en speed around 800\,km\,s$^{-1}$, which is not only much larger than the local bulk speed, but is in excess of the stream's eventual asymptotic speed.

Comparison of these in situ properties with historical UVCS measurements \citep[][]{Cranmer2020b} of PCH flows, shown together in Figure~\ref{fig:fig4_radialevolution} reveal that the radial evolution of this PCH-L stream from a large equatorial coronal hole is highly consistent with the temperature and velocity profiles of protons in PCH flows at solar minimum. A smooth acceleration curve, illustrated here with iso-poly+forcing models \citep[][dashed black curves in Figure~\ref{fig:fig4_radialevolution}B]{Dakeyo2022, Shi2022, Rivera2024,Rivera2025}, whose gradient monotonically decreases with distance above the sonic critical point (here around 2\,$R_\odot$) links these data together and suggests that large equatorial and polar coronal holes behave similarly at all distances, not just in terms of asymptotic properties.

Independent of whether this equatorial stream may be interpreted as globally representative of PCH flows (see further discussion below), we can quantify that 1/3 of the stream's acceleration takes place beyond $10\,R_\odot$ but is almost complete (95~\%) by $60\,R_\odot$. The acceleration may be therefore regarded as gradual and this observation provides strong constraints on where and how quickly energy is deposited into the bulk flow. 

\subsection{Implications of Iso-poly+Forcing profiles}\label{subsec:isopoly-comparison}
The family of iso-poly+forcing profiles shown in Fig \ref{fig:fig4_radialevolution} includes a significant portion of additional forcing supplied empirically (see below). This forcing increases in strength monotonically with asymptotic speed \citep[see][Appendix B]{Badman2025} such that the slowest speed profiles are purely thermal-pressure driven \citep{Halekas2022,Halekas2023,Rivera2025}, the intermediate fast profiles have a boost that was empirically constrained by measurements of an Alfv\'en wave pressure gradient in \citep[][]{Rivera2024} and the fastest profile has a sufficient boost to match the $99th$ percentile of speed measurements from Parker and Solar Orbiter. This latter (99$^{th}$ percentile) profile best matches the stream studied here across density, velocity and temperature. This means that the observed acceleration requires significant work done on the plasma beyond the observed thermal pressure gradients \citep[as expected historically, e.g.][]{Leer1982,Axford1999}. Additionally, the effect here is \textit{gradual} in the sense that a significant portion of this work must occur above the Alfv\'en point (15.2\,$R_\odot$) and remains strongly active at least until the intermediate measurement at $60\,R_\odot$. This is consistent with \citet[][]{Halekas2023,Rivera2024} which showed this additional boost could be explained by Alfv\'enic fluctuation energy present in the stream, and would be expected to be larger in asymptotically faster streams. Quantification and evolution of the Alfv\'enic energy flux, along with other potential contributions to the streams energy budget, will be the subject of future work \citep[][]{Rivera2026_inprep}, but we briefly motivate its importance: By computing the ratio of the average Alfv\'en wave energy density to the average proton kinetic energy density measured at Parker \citep[][equations S4a and S4d]{Rivera2024} which can be written as $U_W/U_{K,p} = (3/2 + 1/M_A)\delta v^2/u^2$. At 10\,$R_\odot$, this fraction is $\sim45\%$ where $\delta v$ was computed on a root mean square basis. Assuming 100\% conversion into kinetic energy of protons, this would contribute $\sim$100\,km\,s$^{-1}$ to the acceleration budget (about 1/3 of what is needed between Parker and 1\,au), which is clearly substantial although further work is needed to establish if it is sufficient to close the energy budget. 

It is also interesting to note that while the model profiles presented here do not include fully self-consistent treatment of the interaction between Alfv\'enic fluctuations and heating in the stream, they qualitatively resemble models that do, especially in terms of the gradual acceleration profile and the turn over in temperature from near isothermal to polytropic cooling  \citep[e.g.,][]{Matthaeus1999, Cranmer2005, Cranmer2007,Verdini2010,Chandran2025,Usmanov2025}.

\subsection{Comparison to Radio Scintillation Observations of PCH Flows}\label{subsec:scintillation}
As mentioned above, the in situ measurements of the acceleration of this stream from a large equatorial coronal hole as close as 10\,$R_\odot$, smoothly connects to Doppler-dimming based measurements of the velocity of PCH acceleration out to $4\,R_\odot$ from UVCS \citep[][]{Cranmer2020a}. However, the 10\,$R_\odot$ in situ measurement here is significantly slower that estimates of the speed determined from radio scintillation measurements at the same distance of PCH wind \citep[][]{Grall1996,Breen1996, Moran1997}, which have been interpreted to show that PCH wind is essentially finished accelerating inside of 10\,$R_\odot$. 

We suggest two possible routes to resolve this discrepancy. The first is that the scintillation estimates are robust which would indicate that the outer coronal acceleration of large coronal holes are qualitatively different if they are equatorial vs. polar, but end up in a near-identical asymptotic state \citep[see, e.g.,][]{Miralles2001}. This would also require a more complicated acceleration profile for polar flows, compared to equatorial flows, in order to smoothly join the UVCS measurements to the scintillation measurements. Such a profile would monotonically flatten up to 4\,$R_\odot$, steepen rapidly before $10\,R_\odot$ and then flatten again rapidly and would be substantially different to the smooth, monotonically flattening curves that link the present stream to UVCS PCH data. 

The second possibility is that the scintillation measurements yield significant overestimates in the 10--15\,$R_\odot$ distance range. This is plausible on the basis that these observations are strongly affected by velocity (and density) fluctuations along the line of sight, which has been suggested to lead to an upward bias in the apparent speed by up to the characteristic wave speed \citep[see e.g. Figure 20 of][, here $V_A$]{Harmon2005}. Indeed, for the present stream, the total $V_R+V_A$ is approximately 1250-1300 km/s at $10 R_\odot$ (Fig \ref{fig:fig1_stream_summary}C), which places it above the maximum scintillation measurements reported in \citep[$\sim$ 1000 km/s][]{Grall1996}. 

We argue that given the first case suggests additional complexity to the acceleration profiles, and that the Parker measurements confirm very large velocity fluctuations and Alfv\'en speed in the stream, the second possibility is more likely. In either case, these new data provide for the first time a ground truth on the bulk flow, Alfven speed and fluctuation amplitude and geometry of such a stream, which can be used to predict how the measured bulk speed here would appear in scintillation measurements which is a promising route to resolve this discrepancy.

If the scintillation results remain robust, there are puzzles to solve: PCH-L wind with sufficiently large coronal holes reaches the same asymptotic state as PCH, but behaves differently en route to get there. This picture therefore requires a physical difference in solar wind acceleration at intermediate stages but a conserved overall energy budget. Low down stream interactions and solar rotation would not be expected to strongly deform things.  A plausible difference could arise from having different Alfv\'en speed profiles, and therefore differences in its gradient), above equatorial and polar coronal holes due to differences in coronal magnetic structure which varies with solar cycle. Differences in wave reflection properties could then lead to different damping rates, while maintaining a consistent energy input at the coronal base \citep[][]{Axford1999}.

If the second possibility is the case and the scintillation measurements are indeed significant overestimates close to the Sun, then we may interpret the present observations of a 750\,km\,s$^{-1}$ stream from a large equatorial coronal hole as robustly representative of true PCH flows observed at solar minimum, which greatly widens the applicability of the constraints and findings of this work. In particular, it suggests all such flows indeed accelerate gradually rather than being fully energized by 10\,$R_\odot$.

\subsection{Fluctuation Evolution and Implications for Switchback Formation}
We examined the evolution of velocity fluctuations in the stream and determine that it is filled with spherically polarized Alfv\'en waves which we characterize with ``Alfv\'en spheres''  at each measurement location. These fluctuations constitute small-angle deflections on the Alfv\'en sphere at $10\,R_\odot$ but develop into large angle ``full reversal switchbacks'' by 60\,$R_\odot$, consistent with recent work showing the spectrum of deflections of magnetic switchbacks grows uniformly with distance and Alfv\'en Mach number \citep[][]{Bowen2025, Goodwill2026, Payne2026}. 

At $10\,R_\odot$, the sphere that the velocity fluctuations trace out has a radius around 800\,km\,s$^{-1}$ (the local Alfv\'en speed). This means that even though the deflections are relatively small angle (20-30$^\circ$) on the sphere and therefore do not appear as impressive in terms of radial velocity spikes, their transverse amplitude is enormous and can even exceed the bulk speed. These fluctuations therefore contain a huge amount of Poynting flux \citep[][]{Woolley2020} compared to the ``full reversal'' state at Solar Orbiter. As mentioned above, this will be quantified in the context of the evolution of the full energy budget in follow up work, but we remark here that it is plausible that these fluctuations, whose energy density is almost half the kinetic energy density at Parker (see above), provide a substantial portion of the extra work needed to explain the continued acceleration of the stream ($\S$~\ref{subsec:isopoly-comparison}, and are large ampltide and with a large enough group velocity ($V_A$) to significantly affect radio scintillation observations (Section~\ref{subsec:scintillation}).

\subsection{Rarity of these Streams in Solar Cycle 25}
Comparison to statistical measurements in Figure \ref{fig:fig4_radialevolution} reveals that such flows in the near-equatorial plane are extremely rare during the range of solar cycle phases probed by Parker to date, with the stream properties lying in the 99th percentile of measurements as a function of radial distance. As noted by \citet{Dakeyo2022} (see their Figure 1, using Parker data through 2022), this appears to be a qualitative difference between the current solar cycle and the previous ones probed close to the Sun ($R > 0.3$\,au) by Helios for which fast wind $>$~600\,km\,s$^{-1}$ was relatively well represented while Parker provided almost no equivalent samples at matching heliocentric distances. 

Given that Parker extended into previously unexamined distance ranges, and this lack of fast wind sampling initially persisted, it is worth noting that the hypothesis that this wasn't a statistical sampling effect but rather a physical indicator of extremely extended acceleration was not ruled out. These new measurements of 450--500\,km\,s$^{-1}$ at 10\,$R_\odot$ confirm that statistical sampling is the much more likely scenario and will continue to be filled in as Parker continues to encounter large coronal holes associated with the equatorial and coronal holes that become more frequent in the declining phase of the solar cycle. It will be interesting to track the extent to which the statistical distribution of velocities measured by Parker and Solar Orbiter over a full solar cycle 25 matches that of Helios, and therefore assess if significant solar cycle differences remain.

\section{Conclusions and Future Work}\label{sec:conclusion}

We summarize these results and discussions by drawing the following major conclusions arising from this study:

\begin{itemize}
    \item This flow measured by Parker at $10\,R_\odot$ is a rare ($> 99th$ percentile by speed) but highly informative and novel event which provides strong constraints on solar wind acceleration theories. The equatorial coronal hole (ECH) source region, and asymptotic properties at 1\,au classify it as polar coronal hole-like (PCH-L) in the sense that: 1) the size of this ECH (7\% of the solar surface) is comparable to solar-minimum polar coronal holes  \citep[6-10\%][]{DeToma2011}; 2) the asymptotic properties are near identical to the homogeneous properties of those flows seen by Ulysses, and at the maximum of high speed streams observed in the ecliptic \citep[for a more global comparison, see ][ and in prep]{Rivera2025_AGU}.
    \item The key highlights of these constraints is that at $10\,R_\odot$, such streams are significantly sub-Alfv\'enic ($M_A\sim0.6$), have reached only 60\% of their asymptotic speed, are near a proton-temperature maximum at or above expected low coronal proton temperatures, and consistent with gradual acceleration that continues to be significant out to at least $\sim60\,R_\odot$. This gradual, monotonic acceleration is consistent with the iso-poly+forcing curves here which qualitatively resemble more physically self-consistent wave+turbulence models \citep[e.g.][]{Verdini2007,Cranmer2007,Verdini2010, Sokolov2013, Lionello2014,Gombosi2018,Chandran2025} and therefore offer qualified support for such physics.
    \item The solar wind speed at $\sim$ 10\,$R_\odot$ (450\,km\,s$^{-1}$) is well below the equivalent speed determined by radio scintillation in PCH flows \citep[][]{Grall1996,Breen1996,Moran1997}, which suggested rapid acceleration which is effectively finished by $10\,R_\odot$. The resolution to this is either a qualitative, and physically interesting difference in the low- and mid-coronal acceleration profile between large polar and equatorial coronal holes (with asymptotic properties identical), or a systematic overestimation of the bulk speeds close to the Sun from radio scintillation due to their convolving of the bulk and wave velocitiyes \citep[][]{Harmon2005}. Either outcome would be interesting and can be performed by forward modeling scintillation observations for this or similar streams with the known ground truth bulk, fluctuation flows and Alfv\'en speeds provided by in situ data \citep[as suggested by ][]{Grall1996}.
    \item The evolution of velocity fluctuations in the stream is highly informative on the development of magnetic switchbacks in the solar wind. Specifically, the fluctuations of the 3D velocity vector are observed to be spherically polarized but smaller (20-30$^\circ$) angle at $10\,R_\odot$ such that their transverse fluctuation components dominate and are comparable or even larger than the bulk flow. However, by $60\,R_\odot$, they are large angle but traveling on a much smaller sphere in velocity space (radius $\sim$ $V_A$) and resemble the full reversal switchbacks observed by Parker Solar Probe in its first encounters. 
\end{itemize}

Moving forward, this stream (and others like it which are becoming more common in the declining solar cycle phase) warrants numerous future investigations.

The source(s) of this specific fast wind stream, and others like it, were subject to coordinated remote-sensing campaigns with Solar Orbiter and near-Earth satelites (e.g. Hinode and IRIS) in 2025 and 2026. In future, these observations will provide additional insight from low down at the source as to how the solar wind forms and how energy is injected to fuel its acceleration. Additionally, recent remote sensing observatories have and will continue to make available new constraints in the low corona on the acceleration profiles of different types of solar wind, for example see recent results odvecting density structures from PROBA-3 \citep[][]{Zhukov2026} and planned flow tracking products from PUNCH \citep[][]{Deforest2026}, as well as coronal hole flow tracking from Solar Orbiter/METIS \citep[][]{Giordano2025}.

The discrepancy between the implied acceleration curve for this PCH-L flow with that suggested by radio scintillation should be resolved by using the measurements of the bulk ($V_R$) and fluctuation ($d\textbf{V}$) velocity and density fluctuations ($dN$) at Parker Solar Probe to forward model a scintillation observation of this stream and determine if its true speed is captured or overestimated. 

The energy budget of this and similar streams should be computed at these three measurement locations to separate and quantify the contributions to the acceleration from thermal pressures, ambipolar potential, reduction in Alfv\'enic Poynting flux (estimated in this work as 45\% of the kinetic energy flux at 10\,$R_\odot$), and potential non-negligible contributions from Helium whose abundance and mass flux contribution is known to be highest in fast streams.

The turbulence properties of this stream, which has recently been shown to be highly Alfv\'enic \citep[][]{Bandyopadhyay2026_inrev}, should be studied as a function of radial evolution to investigate the development and evolution of different driving and dissipation processes, although the lack of Solar Orbiter MAG data at the peak of the stream unfortunately precludes independent characterization of dV and dB at the intermediate measurement point.

Repeat examples of this kind of stream should be obtained as often as measured to add statistical support to these findings. Although, given the homogeneity of Ulysses solar minimum PCH flow, we expect these PCH-L flows to be similarly repeatable, independent evidence is needed to fully confirm this. In 2025, the large-scale structure of the Sun's magnetic field was influenced by the strong nesting of active regions in 2024, that resulted in an inclined dipolar field that persisted for several months \citep[][and in prep.]{Finley2025, finley2026arxiv, Rivera2025_AGU} with this same stream observed over multiple Carrington rotations. This already provides multiple opportunities to study how different large-scale field topologies affect the fast wind acceleration and heating.

One particular aspect of such complexity and variation not addressed in this work, is the clear substructure in the stream which we reduce here to simple medians and distributions across the stream. In particular we note the presence of at least three distinct switchback patches or microstreams in the peak of the stream at Parker \citep[][]{Jaffarove2026} which are plausibly identifiable in the later crossings of the stream but somewhat smoothed out. The dynamics internal to the stream between these substructures and how well their evolution can be modeled and tied to coronal source structure would provide a more complete and detailed accounting for this evolution. This could further explain the transition from ubiquitous strong (100s of km/s) velocity contrasts/striations observed close to the Sun by remote sensing methods \citep[][]{DeForest2018} to the more homogeneous structure observed in PCH flows beyond 1\,au \citep[][]{McComas2000}, alongside addressing the connection \citep[][and in prep]{Ahmed2025_AGU} between switchback patches \citep[][]{Bale2021,Fargette2021,Shi2022} and what have historically been termed microstreams farther from the Sun \citep[e.g.][]{Neugebauer1995,Horbury2023}. Additionally, the latitude structure of this stream \citep[e.g.][and in prep]{Planet2024_AGU} may be probed through use of the inbound and outbound crossings of the stream by Parker, as well as the cut through the stream by STEREO A \citep[][]{Kaiser2008} at 1\,au a few days prior to \textit{Wind}.

Finally, it is worth noting that the polar coronal holes are again starting to form towards the solar rotational poles (i.e., towards higher latitudes) at the time of writing. For example, in Figure \ref{fig:fig1_stream_summary}A, the largest portion of the PCH-L coronal hole studied here is towards the South. Future Solar Orbiter data, which is increasing its orbital inclination, may soon measure such flows directly from the largest portion of coronal holes and give further clues on unresolved differences between ECH/PCH-L and PCH flows at they transform into each other.

\begin{acknowledgments}
Parker Solar Probe was designed, built, and is now operated by the Johns Hopkins Applied Physics Laboratory as part of NASA’s Living with a Star (LWS) program (contract NNN06AA01C). Support from the LWS management and technical team has played a critical role in the success of the Parker Solar Probe mission. 

Solar Orbiter is a mission of international cooperation between ESA and NASA, operated by ESA. Solar Orbiter SWA data were derived from scientific sensors that were designed and created and are operated under funding provided by numerous contracts from UKSA, STFC, the Italian Space Agency, CNES, the French National Centre for Scientiﬁc Research, the Czech contribution to the ESA PRODEX program, and NASA.  Solar Orbiter SWA work at the UCL/Mullard Space Science Laboratory is currently funded by STFC (grant Nos. ST/W001  Solar Orbiter magnetometer operations are funded by the UK Space Agency (grant UKRI943).004/1 and ST/X/002152/1).

"SDO data are supplied courtesy of the SDO/HMI and SDO/AIA consortia. SDO is the first mission to be launched for NASA's Living With a Star (LWS) Program.

 S.T.B., Y.J.R, S.B.D, M.T., K.P., L.Y.A, M.L.S., F.F. and T.N. were partially supported by Parker Solar Probe project through the SAO/SWEAP subcontract 975569.

 S.D.B. acknowledges support from the Royal Society Wolfson Visiting Fellowship program.

 T.E. acknowledges funding support from The Chuck Lorre Big Bang Theory Graduate Fellowship.

 R.B., S.P.G., and W.H.M. are partially supported by the Parker Solar Probe Project through  subcontract SUB0000165 at University of Delaware from Princeton University's ISOIS team. R.B. acknowledges support from NSF award number 2347952 (FDSS Track 1: Expansion and Integration of Space Physics at the University of Delaware)

T.H. is supported by STFC grant ST/W001071/1. 

AJF acknowledges support through the European Space Agency (ESA) Research Fellowship in Space Science.

\end{acknowledgments}

\begin{contribution}


\end{contribution}

\software{astropy \citep{astropy:2013,astropy:2018,astropy:2022}, 
pyspedas \citep{Angelopoulous2020}, 
sunkit-magex \citep[][]{Stansby2020},
SunPy \citep{Sunpy2020}, ParkerSolarWind \citep[][]{Badman2023} 
}

\appendix

\section{In Situ Measurement Details}
\label{appendix:instruments}

Here, we provide an accounting of all non-trivial measurement details for the results presented in this work. In particular, our focus is on obtaining proton density, velocity and temperatures and electron temperatures and we have neglected the alpha particle population in general.

\subsection{Magnetic Field}\label{appendix:B}

Magnetic field data are some of the simpler measurements to interpret across all spacecraft. Data for Parker comes from the 4 samples/cycle FIELDS \citep[][]{Bale2016} data product. 

Data for Solar Orbiter comes from the MAG instrument \citep[][]{Horbury2020}, although we note that due to a spacecraft anomaly at the time of the peak of this stream which means no magnetic field data is available for the exact timerange of this interval at Solar Orbiter.

For \emph{Wind} \citep[][]{wilsoniii21a}, magnetic field data comes from the Magnetic Field investigation \citep[MFI][]{lepping95, windmfi3s21a}.

\subsection{Plasma Density}\label{appendix:N}

For Parker, we utilize the quasithermal noise (QTN) data product derived from the FIELDS Radio Frequency Spectrometer \citep[RFS;][]{Pulupa2017} low frequency range (LFR) spectra, as described in \citet{Moncuquet2020}. This provides an accurate measurement of the electron density when the spacecraft is close enough to the Sun that the plasma frequency is in the frequency range of LFR and provided there are no large radio events overlapping the same frequency range. These conditions are both met for the stream studied in this work. Direct measurements of the proton density are not available from the Solar Probe Cup \citep[][]{Case2020} at these heliocentric distances, and measurements of the density from Solar Probe Analyzer-ions \citep[SPAN-i;][]{Livi2022} are strongly perturbed by Alfv\'enic velocity spikes, meaning additional processing is required to interpret them, meaning QTN is the most straightforward and accurate plasma density measurement available.

We interpret the electron density measured here as approximately the same as the proton density so that it can be directly compared to the proton densities, which are the more robust measurements at the other spacecraft used in this work. This assumption is worth briefly quantifying since fast wind streams typically have the highest alpha abundances observed in the solar wind \citep[$N_\alpha/N_p \sim$4\%][]{Alterman2019} apart from during CMEs \citep[][]{Johnson2024}. For a safe margin, we consider a worst case  8\% abundance. In this case, we get $N_p + 2N_{\alpha} = N_p(1+2N_\alpha/N_p)=N_e$ which yields an electron density 16\% larger than protons. With this exaggerated abundance, the potential shift is slightly smaller than the interquartile range of the Parker Ne across the stream; we therefore consider it a small effect. 


For Solar Orbiter, we utilize the proton density moment from the Proton and Alpha Sensor \citep[;PAS][]{Owen2020,Louarn2021}. The PAS VDFs initially are the joint distribution of proton and alpha particles. To extract the proton moments, a peak tracking procedure has been implemented in the L2 data product to exclude the alpha particles. For the data used in this paper, the two species are well separated and thus the resulting moments are approximately only reflecting the proton component of the solar wind. 

While this measurement is much more robust to field of view effects as compared to SPAN-i, due to the particularly large velocity spikes in this stream, we still observe some correlated density dropouts which we therefore consider unphysical (cf. Fig. \ref{fig:fig2_stream_properties} E \& H). We therefore condition the density measurements on the velocity spikes (V$R$ < 1.1 Median($V_R$) with the median computed on a 1 hour sliding window) to build the distribution to produce the box plot shown in Fig. \ref{fig:fig4_radialevolution}A.

At \emph{Wind}, we use the Solar Wind Experiment \citep[SWE; ][]{Ogilvie1995} nonlinear proton fits. Here, the measurement is most directly interpretable as a true measurement of $N_p$, well-differentiated from alphas. 

We note that examining Figure \ref{fig:fig4_radialevolution}A, the Parker median density is slightly higher than the mass flux curve closest to the Solar Orbiter and \emph{Wind} proton densities. It is therefore plausible that accounting for the effect of finite alpha abundances on quasineutrality would lead to a decreases in the median density for Parker, and match mass flux conservation even more closely.

\subsection{Proton Velocity and Fluctuations}\label{appendix:V}

For Parker, we use the SPAN-i velocity vector moments in the RTN frame. These are generally the most robust SPAN-i data product. We do not filter these moments based on field of view, but instead illustrate in Figure \ref{fig:fig3_switchbacks} that finite field of view effects clearly manifest by restricting the portion of velocity space which SPAN-i can make measurements of. In particular, Figure \ref{fig:fig3_switchbacks}A-C shows that during this stream, large +$V_T$ velocity deflections are not well measured, but the background flow in the radial direction is well measured such that the bulk speed is well-determined, and that there are sufficient deflections in the $\pm$N and -T directions to adequately sample the velocity fluctuations.

For Solar Orbiter, we use the PAS velocity vector moment which is provided in RTN coordinates. By comparing and contrasting these measurements to those of SPAN-i in Figure \ref{fig:fig3_switchbacks}, we confirm the measurement population is not significantly impeded by field of view effects.

For \emph{Wind} we again use SWE non-linear fits which provide the velocity vector moment in Geocentric Earth-Ecliptic (GSE) coordinates, which we convert to RTN through the matching : GSE +X = -R, GSE +Y = -T, GSE +Z = +N.

\subsection{Proton Temperature}\label{appendix:Tp}

For Parker, we use the SPAN-i spacecraft frame temperature tensor. We apply the same FOV as described above for the velocity moments. Next, we follow the procedure describe in \citet[][]{Badman2025} Appendix A to diagonalize the tensor in the magnetic field direction and compute $T_\perp$ and $T_\parallel$ in a way that rejects the tensor components most strongly impeded by the SPAN-i FOV affects. The isotropic proton temperature is then computed as ($2/3T_\perp + 1/3T_\parallel$). Compared to the tensor trace, this correction constitutes a slight increase in the isotropic temperature because the typical impact of the FOV issue is to reduce the apparent value of one of the $T_\perp$ temperature components due to the VDF being truncated in that direction. While this transformation sometimes blows up due to the magnetic field vector being parallel to the poor measurement direction, these instances are filtered out as anomalously high values.

For Solar Orbiter, we use the trace of the temperature tensor moment, which gives the isotropic temperature ($2/3T_\perp + 1/3T_\parallel$). To avoid changing FOV effects on the edge of the distribution we apply the same velocity spike condition as for the PAS density on the temperature moments which are trusted.

 For \emph{Wind}, we again use the SWE non-linear proton fits which provide a thermal speed ($w_{th,p}$) which we convert to units of temperature as $k_B T_p = \frac{1}{2}m_pw_{th,p}^2$. The SWE measurements over a substantial fraction of a spacecraft spin period (3 seconds) are used to discriminate the perpendicular and parallel directions and provide a scalar temperature equivalent to the tensor trace provided by the other spacecrafts' electrostatic analyzers.
 
\subsection{Electron Temperature}\label{appendix:Te}

At Parker, we utilize core fits of the SPAN-e \citep[][]{Whittlesey2020} electron velocity distribution function (eVDF), computed using the methodology from \citet{Halekas2020}. 


At Solar Orbiter, electron temperatures are obtained from core fits to eVDFs measured by SWA/EAS \citep{Owen2020}, using the fitting method of \citet{wu2026}.
The energy channels of eVDFs are first corrected by subtracting the spacecraft potential $\phi_{sc}$ provided by RPW \citep{maksimovic_rpw}, and the fit is restricted to corrected energies above 14 eV to exclude spacecraft-emitted electrons \citep{stverak2025}.
An uncertainty in $\phi_{sc}$ shifts the energy channels as a whole, so it mainly displaces the low-energy boundary of the fit by a few electronvolts. The electron temperature, set by the slope of the logarithm of the distribution versus energy, is largely insensitive to such uncertainties \citep{genot2004}. 
During 30-31 March 2025, when magnetometer data are unavailable, a drifting anisotropic Maxwellian is fitted in the spacecraft reference frame, and the electron temperature is taken as one third of the trace of the fitted temperature tensor. As the trace is rotation-invariant, this estimate requires no knowledge of the field direction. 
On adjacent days where both approaches can be applied, the two agree well.

At \emph{Wind}, no cross-validation with the spacecraft potential was possible which made the interpretaion of SWE \citep[][]{ogilvie95, windswestrahlh521a} electron temperature moments unreliable. Instead, we provide the statistical range of electron temperatures measured by 3DP \citep[][]{lin95a} for fast wind (Vsw $>$ 500\,km\,s$^{-1}$) at 1\,au \citep[][]{Wilson2023, Wilson2026}, as well as an estimate of typical values in PCH-L wind (Vsw $>$ 800\,km\,s$^{-1}$) also from \citep[][]{Wilson2026}, and in true PCH wind from Ulysses data \citep[][]{Phillips1996}. These latter two measurements are important because there remains systematic variation (anticorrelation) between $V_{SW}$ and $T_e$ for velocities variation above 500\,km\,s$^{-1}$ which tells us that, the expected electron temperature in 800\,km\,s$^{-1}$ wind is in the lower percentile range of all wind faster that 500\,km\,s$^{-1}$ at 1\,au.  The 3DP electron data from \citet[][]{Wilson2026} were calibrated by first determining and accounting for the spacecraft potential \citep[][]{wilsoniii23a, wilsoniii23b} then altering the anode corrections until the total electron density determined from the the upper hybrid line matches the zeroth velocity moment from numerically integrating the velocity distribution \citep[][]{martinovic20a}.  A similar approach was used for the electron moments in \citet[][]{Wilson2023} except that the spacecraft potential was a free parameter that would vary until the integrated number density matched the total electron density determined from the the upper hybrid line.  The reason for the difference is the minimum energy of the 3DP electron detector was too high for much of the time frame prior to 2005 to directly observe the photoelectrons used to infer the spacecraft potential as was done by \citet[][]{wilsoniii23a}.




\bibliography{sample7}

\begin{thebibliography}{}
\expandafter\ifx\csname natexlab\endcsname\relax\def\natexlab#1{#1}\fi
\providecommand{\url}[1]{\href{#1}{#1}}
\providecommand{\dodoi}[1]{doi:~\href{http://doi.org/#1}{\nolinkurl{#1}}}
\providecommand{\doeprint}[1]{\href{http://ascl.net/#1}{\nolinkurl{http://ascl.net/#1}}}
\providecommand{\doarXiv}[1]{\href{https://arxiv.org/abs/#1}{\nolinkurl{https://arxiv.org/abs/#1}}}

\bibitem[{L. {Ahmed} {et~al.}(2025){Ahmed}, {Stevens}, {Badman}, {Rivera}, {Paulson}, \& {Das}}]{Ahmed2025_AGU}
{Ahmed}, L., {Stevens}, M.~L., {Badman}, S.~T., {et~al.} 2025, in AGU Fall Meeting Abstracts, Vol. 2025, AGU Fall Meeting Abstracts, SH32A--08

\bibitem[{B.~L. {Alterman} \& J.~C. {Kasper}(2019){Alterman} \& {Kasper}}]{Alterman2019}
{Alterman}, B.~L., \& {Kasper}, J.~C. 2019, \bibinfo{title}{{Helium Variation across Two Solar Cycles Reveals a Speed-dependent Phase Lag},} \apjl, 879, L6, \dodoi{10.3847/2041-8213/ab2391}

\bibitem[{B.~L. {Alterman} {et~al.}(2025){Alterman}, {Rivera}, {Lepri}, \& {Raines}}]{Alterman2025}
{Alterman}, B.~L., {Rivera}, Y.~J., {Lepri}, S.~T., \& {Raines}, J.~M. 2025, \bibinfo{title}{{The transition from slow to fast wind as observed in composition observations},} \aap, 694, A265, \dodoi{10.1051/0004-6361/202451550}

\bibitem[{M.~D. {Altschuler} \& G. {Newkirk}(1969){Altschuler} \& {Newkirk}}]{altschuler1969_pfss}
{Altschuler}, M.~D., \& {Newkirk}, Jr., G. 1969, \bibinfo{title}{{Magnetic Fields and the Structure of the Solar Corona. I: Methods of Calculating Coronal Fields},} \solphys, 9, 131, \dodoi{10.1007/BF00145734}

\bibitem[{V. Angelopoulos {et~al.}(2019)Angelopoulos, Cruce, Drozdov, Grimes, Hatzigeorgiu, King, Larson, Lewis, McTiernan, Roberts, Russell, Hori, Kasahara, Kumamoto, Matsuoka, Miyashita, Miyoshi, Shinohara, Teramoto, Faden, Halford, McCarthy, Millan, Sample, Smith, Woodger, Masson, Narock, Asamura, Chang, Chiang, Kazama, Keika, Matsuda, Segawa, Seki, Shoji, Tam, Umemura, Wang, Wang, Redmon, Rodriguez, Singer, Vandegriff, Abe, Nose, Shinbori, Tanaka, UeNo, Andersson, Dunn, Fowler, Halekas, Hara, Harada, Lee, Lillis, Mitchell, Argall, Bromund, Burch, Cohen, Galloy, Giles, Jaynes, Le~Contel, Oka, Phan, Walsh, Westlake, Wilder, Bale, Livi, Pulupa, Whittlesey, DeWolfe, Harter, Lucas, Auster, Bonnell, Cully, Donovan, Ergun, Frey, Jackel, Keiling, Korth, McFadden, Nishimura, Plaschke, Robert, Turner, Weygand, Candey, Johnson, Kovalick, Liu, McGuire, Breneman, Kersten, \& Schroeder}]{Angelopoulous2020}
Angelopoulos, V., Cruce, P., Drozdov, A., {et~al.} 2019, \bibinfo{title}{The Space Physics Environment Data Analysis System (SPEDAS),} Space Science Reviews, 215, 9, \dodoi{10.1007/s11214-018-0576-4}

\bibitem[{C.~N. {Arge} {et~al.}(2010){Arge}, {Henney}, {Koller}, {Compeau}, {Young}, {MacKenzie}, {Fay}, \& {Harvey}}]{Arge2010}
{Arge}, C.~N., {Henney}, C.~J., {Koller}, J., {et~al.} 2010, in American Institute of Physics Conference Series, Vol. 1216, Twelfth International Solar Wind Conference, ed. M.~{Maksimovic}, K.~{Issautier}, N.~{Meyer-Vernet}, M.~{Moncuquet}, \& F.~{Pantellini} (AIP), 343--346, \dodoi{10.1063/1.3395870}

\bibitem[{C.~N. {Arge} {et~al.}(2003){Arge}, {Odstrcil}, {Pizzo}, \& {Mayer}}]{Arge2003}
{Arge}, C.~N., {Odstrcil}, D., {Pizzo}, V.~J., \& {Mayer}, L.~R. 2003, in American Institute of Physics Conference Series, Vol. 679, Solar Wind Ten, ed. M.~{Velli}, R.~{Bruno}, F.~{Malara}, \& B.~{Bucci} (AIP), 190--193, \dodoi{10.1063/1.1618574}

\bibitem[{ {Astropy Collaboration} {et~al.}(2013){Astropy Collaboration}, {Robitaille}, {Tollerud}, {Greenfield}, {Droettboom}, {Bray}, {Aldcroft}, {Davis}, {Ginsburg}, {Price-Whelan}, {Kerzendorf}, {Conley}, {Crighton}, {Barbary}, {Muna}, {Ferguson}, {Grollier}, {Parikh}, {Nair}, {Unther}, {Deil}, {Woillez}, {Conseil}, {Kramer}, {Turner}, {Singer}, {Fox}, {Weaver}, {Zabalza}, {Edwards}, {Azalee Bostroem}, {Burke}, {Casey}, {Crawford}, {Dencheva}, {Ely}, {Jenness}, {Labrie}, {Lim}, {Pierfederici}, {Pontzen}, {Ptak}, {Refsdal}, {Servillat}, \& {Streicher}}]{astropy:2013}
{Astropy Collaboration}, {Robitaille}, T.~P., {Tollerud}, E.~J., {et~al.} 2013, \bibinfo{title}{{Astropy: A community Python package for astronomy},} \aap, 558, A33, \dodoi{10.1051/0004-6361/201322068}

\bibitem[{ {Astropy Collaboration} {et~al.}(2018){Astropy Collaboration}, {Price-Whelan}, {Sip{\H{o}}cz}, {G{\"u}nther}, {Lim}, {Crawford}, {Conseil}, {Shupe}, {Craig}, {Dencheva}, {Ginsburg}, {VanderPlas}, {Bradley}, {P{\'e}rez-Su{\'a}rez}, {de Val-Borro}, {Aldcroft}, {Cruz}, {Robitaille}, {Tollerud}, {Ardelean}, {Babej}, {Bach}, {Bachetti}, {Bakanov}, {Bamford}, {Barentsen}, {Barmby}, {Baumbach}, {Berry}, {Biscani}, {Boquien}, {Bostroem}, {Bouma}, {Brammer}, {Bray}, {Breytenbach}, {Buddelmeijer}, {Burke}, {Calderone}, {Cano Rodr{\'\i}guez}, {Cara}, {Cardoso}, {Cheedella}, {Copin}, {Corrales}, {Crichton}, {D'Avella}, {Deil}, {Depagne}, {Dietrich}, {Donath}, {Droettboom}, {Earl}, {Erben}, {Fabbro}, {Ferreira}, {Finethy}, {Fox}, {Garrison}, {Gibbons}, {Goldstein}, {Gommers}, {Greco}, {Greenfield}, {Groener}, {Grollier}, {Hagen}, {Hirst}, {Homeier}, {Horton}, {Hosseinzadeh}, {Hu}, {Hunkeler}, {Ivezi{\'c}}, {Jain}, {Jenness}, {Kanarek}, {Kendrew}, {Kern}, {Kerzendorf}, {Khvalko}, {King}, {Kirkby}, {Kulkarni},
  {Kumar}, {Lee}, {Lenz}, {Littlefair}, {Ma}, {Macleod}, {Mastropietro}, {McCully}, {Montagnac}, {Morris}, {Mueller}, {Mumford}, {Muna}, {Murphy}, {Nelson}, {Nguyen}, {Ninan}, {N{\"o}the}, {Ogaz}, {Oh}, {Parejko}, {Parley}, {Pascual}, {Patil}, {Patil}, {Plunkett}, {Prochaska}, {Rastogi}, {Reddy Janga}, {Sabater}, {Sakurikar}, {Seifert}, {Sherbert}, {Sherwood-Taylor}, {Shih}, {Sick}, {Silbiger}, {Singanamalla}, {Singer}, {Sladen}, {Sooley}, {Sornarajah}, {Streicher}, {Teuben}, {Thomas}, {Tremblay}, {Turner}, {Terr{\'o}n}, {van Kerkwijk}, {de la Vega}, {Watkins}, {Weaver}, {Whitmore}, {Woillez}, {Zabalza}, \& {Astropy Contributors}}]{astropy:2018}
{Astropy Collaboration}, {Price-Whelan}, A.~M., {Sip{\H{o}}cz}, B.~M., {et~al.} 2018, \bibinfo{title}{{The Astropy Project: Building an Open-science Project and Status of the v2.0 Core Package},} \aj, 156, 123, \dodoi{10.3847/1538-3881/aabc4f}

\bibitem[{ {Astropy Collaboration} {et~al.}(2022){Astropy Collaboration}, {Price-Whelan}, {Lim}, {Earl}, {Starkman}, {Bradley}, {Shupe}, {Patil}, {Corrales}, {Brasseur}, {N{\"o}the}, {Donath}, {Tollerud}, {Morris}, {Ginsburg}, {Vaher}, {Weaver}, {Tocknell}, {Jamieson}, {van Kerkwijk}, {Robitaille}, {Merry}, {Bachetti}, {G{\"u}nther}, {Aldcroft}, {Alvarado-Montes}, {Archibald}, {B{\'o}di}, {Bapat}, {Barentsen}, {Baz{\'a}n}, {Biswas}, {Boquien}, {Burke}, {Cara}, {Cara}, {Conroy}, {Conseil}, {Craig}, {Cross}, {Cruz}, {D'Eugenio}, {Dencheva}, {Devillepoix}, {Dietrich}, {Eigenbrot}, {Erben}, {Ferreira}, {Foreman-Mackey}, {Fox}, {Freij}, {Garg}, {Geda}, {Glattly}, {Gondhalekar}, {Gordon}, {Grant}, {Greenfield}, {Groener}, {Guest}, {Gurovich}, {Handberg}, {Hart}, {Hatfield-Dodds}, {Homeier}, {Hosseinzadeh}, {Jenness}, {Jones}, {Joseph}, {Kalmbach}, {Karamehmetoglu}, {Ka{\l}uszy{\'n}ski}, {Kelley}, {Kern}, {Kerzendorf}, {Koch}, {Kulumani}, {Lee}, {Ly}, {Ma}, {MacBride}, {Maljaars}, {Muna}, {Murphy}, {Norman},
  {O'Steen}, {Oman}, {Pacifici}, {Pascual}, {Pascual-Granado}, {Patil}, {Perren}, {Pickering}, {Rastogi}, {Roulston}, {Ryan}, {Rykoff}, {Sabater}, {Sakurikar}, {Salgado}, {Sanghi}, {Saunders}, {Savchenko}, {Schwardt}, {Seifert-Eckert}, {Shih}, {Jain}, {Shukla}, {Sick}, {Simpson}, {Singanamalla}, {Singer}, {Singhal}, {Sinha}, {Sip{\H{o}}cz}, {Spitler}, {Stansby}, {Streicher}, {{\v{S}}umak}, {Swinbank}, {Taranu}, {Tewary}, {Tremblay}, {de Val-Borro}, {Van Kooten}, {Vasovi{\'c}}, {Verma}, {de Miranda Cardoso}, {Williams}, {Wilson}, {Winkel}, {Wood-Vasey}, {Xue}, {Yoachim}, {Zhang}, {Zonca}, \& {Astropy Project Contributors}}]{astropy:2022}
{Astropy Collaboration}, {Price-Whelan}, A.~M., {Lim}, P.~L., {et~al.} 2022, \bibinfo{title}{{The Astropy Project: Sustaining and Growing a Community-oriented Open-source Project and the Latest Major Release (v5.0) of the Core Package},} \apj, 935, 167, \dodoi{10.3847/1538-4357/ac7c74}

\bibitem[{W.~I. {Axford} {et~al.}(1999){Axford}, {McKenzie}, {Sukhorukova}, {Banaszkiewicz}, {Czechowski}, \& {Ratkiewicz}}]{Axford1999}
{Axford}, W.~I., {McKenzie}, J.~F., {Sukhorukova}, G.~V., {et~al.} 1999, \bibinfo{title}{{Acceleration of the High Speed Solar Wind in Coronal Holes},} \ssr, 87, 25, \dodoi{10.1023/A:1005197529250}

\bibitem[{S. {Badman}(2023){Badman}}]{Badman2023}
{Badman}, S. 2023, \bibinfo{title}{{STBadman/ParkerSolarWind: Release to Zenodo},}, initial release zenodo Zenodo, \dodoi{10.5281/zenodo.10257870}

\bibitem[{S.~T. {Badman} {et~al.}(2021){Badman}, {Bale}, {Rouillard}, {Bowen}, {Bonnell}, {Goetz}, {Harvey}, {MacDowall}, {Malaspina}, \& {Pulupa}}]{Badman2021}
{Badman}, S.~T., {Bale}, S.~D., {Rouillard}, A.~P., {et~al.} 2021, \bibinfo{title}{{Measurement of the open magnetic flux in the inner heliosphere down to 0.13 AU},} \aap, 650, A18, \dodoi{10.1051/0004-6361/202039407}

\bibitem[{S.~T. {Badman} {et~al.}(2023){Badman}, {Stevens}, {Paulson}, {Rivera}, {Niembro Hernandez}, {Verniero}, {Livi}, {Larson}, {Kasper}, {Dakeyo}, \& {Riley}}]{Badman2023_AGU}
{Badman}, S.~T., {Stevens}, M.~L., {Paulson}, K.~W., {et~al.} 2023, in AGU Fall Meeting Abstracts, Vol. 2023, AGU Fall Meeting Abstracts, SH33B--01

\bibitem[{S.~T. {Badman} {et~al.}(2025){Badman}, {Stevens}, {Bale}, {Rivera}, {Klein}, {Niembro}, {Chhiber}, {Rahmati}, {Whittlesey}, {Livi}, {Larson}, {Owen}, {Paulson}, {Horbury}, {Morris}, {O'Brien}, {Dakeyo}, {Verniero}, {Martinovic}, {Pulupa}, \& {Fraschetti}}]{Badman2025}
{Badman}, S.~T., {Stevens}, M.~L., {Bale}, S.~D., {et~al.} 2025, \bibinfo{title}{{Multispacecraft Measurements of the Evolving Geometry of the Solar Alfv{\'e}n Surface over Half a Solar Cycle},} \apjl, 995, L37, \dodoi{10.3847/2041-8213/ae0e5c}

\bibitem[{S.~T. {Badman} {et~al.}(2026){Badman}, {Fargette}, {Matteini}, {Agapitov}, {Akhavan-Tafti}, {Bale}, {Bharati Das}, {Bizien}, {Bowen}, {Dudok de Wit}, {Froment}, {Horbury}, {Huang}, {Jagarlamudi}, {Larosa}, {Madjarska}, {Panasenco}, {Pariat}, {Raouafi}, {Rouillard}, {Ruffolo}, {Sioulas}, {Soni}, {Sorriso-Valvo}, {Suen}, {Velli}, \& {Verniero}}]{Badman2026}
{Badman}, S.~T., {Fargette}, N., {Matteini}, L., {et~al.} 2026, \bibinfo{title}{{Properties of Magnetic Switchbacks in the Near-Sun Solar Wind},} \ssr, 222, 14, \dodoi{10.1007/s11214-026-01267-w}

\bibitem[{S.~D. {Bale} {et~al.}(2016){Bale}, {Goetz}, {Harvey}, {Turin}, {Bonnell}, {Dudok de Wit}, {Ergun}, {MacDowall}, {Pulupa}, {Andre}, {Bolton}, {Bougeret}, {Bowen}, {Burgess}, {Cattell}, {Chandran}, {Chaston}, {Chen}, {Choi}, {Connerney}, {Cranmer}, {Diaz-Aguado}, {Donakowski}, {Drake}, {Farrell}, {Fergeau}, {Fermin}, {Fischer}, {Fox}, {Glaser}, {Goldstein}, {Gordon}, {Hanson}, {Harris}, {Hayes}, {Hinze}, {Hollweg}, {Horbury}, {Howard}, {Hoxie}, {Jannet}, {Karlsson}, {Kasper}, {Kellogg}, {Kien}, {Klimchuk}, {Krasnoselskikh}, {Krucker}, {Lynch}, {Maksimovic}, {Malaspina}, {Marker}, {Martin}, {Martinez-Oliveros}, {McCauley}, {McComas}, {McDonald}, {Meyer-Vernet}, {Moncuquet}, {Monson}, {Mozer}, {Murphy}, {Odom}, {Oliverson}, {Olson}, {Parker}, {Pankow}, {Phan}, {Quataert}, {Quinn}, {Ruplin}, {Salem}, {Seitz}, {Sheppard}, {Siy}, {Stevens}, {Summers}, {Szabo}, {Timofeeva}, {Vaivads}, {Velli}, {Yehle}, {Werthimer}, \& {Wygant}}]{Bale2016}
{Bale}, S.~D., {Goetz}, K., {Harvey}, P.~R., {et~al.} 2016, \bibinfo{title}{{The FIELDS Instrument Suite for Solar Probe Plus. Measuring the Coronal Plasma and Magnetic Field, Plasma Waves and Turbulence, and Radio Signatures of Solar Transients},} \ssr, 204, 49, \dodoi{10.1007/s11214-016-0244-5}

\bibitem[{S.~D. {Bale} {et~al.}(2019){Bale}, {Badman}, {Bonnell}, {Bowen}, {Burgess}, {Case}, {Cattell}, {Chandran}, {Chaston}, {Chen}, {Drake}, {de Wit}, {Eastwood}, {Ergun}, {Farrell}, {Fong}, {Goetz}, {Goldstein}, {Goodrich}, {Harvey}, {Horbury}, {Howes}, {Kasper}, {Kellogg}, {Klimchuk}, {Korreck}, {Krasnoselskikh}, {Krucker}, {Laker}, {Larson}, {MacDowall}, {Maksimovic}, {Malaspina}, {Martinez-Oliveros}, {McComas}, {Meyer-Vernet}, {Moncuquet}, {Mozer}, {Phan}, {Pulupa}, {Raouafi}, {Salem}, {Stansby}, {Stevens}, {Szabo}, {Velli}, {Woolley}, \& {Wygant}}]{Bale2019}
{Bale}, S.~D., {Badman}, S.~T., {Bonnell}, J.~W., {et~al.} 2019, \bibinfo{title}{{Highly structured slow solar wind emerging from an equatorial coronal hole},} \nat, 576, 237, \dodoi{10.1038/s41586-019-1818-7}

\bibitem[{S.~D. {Bale} {et~al.}(2021){Bale}, {Horbury}, {Velli}, {Desai}, {Halekas}, {McManus}, {Panasenco}, {Badman}, {Bowen}, {Chandran}, {Drake}, {Kasper}, {Laker}, {Mallet}, {Matteini}, {Phan}, {Raouafi}, {Squire}, {Woodham}, \& {Woolley}}]{Bale2021}
{Bale}, S.~D., {Horbury}, T.~S., {Velli}, M., {et~al.} 2021, \bibinfo{title}{{A Solar Source of Alfv{\'e}nic Magnetic Field Switchbacks: In Situ Remnants of Magnetic Funnels on Supergranulation Scales},} \apj, 923, 174, \dodoi{10.3847/1538-4357/ac2d8c}

\bibitem[{R. Bandyopadhyay {et~al.}(2026)Bandyopadhyay, Gautam, Matthaeus, Rivera, \& Badman}]{Bandyopadhyay2026_inrev}
Bandyopadhyay, R., Gautam, S.~P., Matthaeus, W.~H., Rivera, Y.~J., \& Badman, S.~T. 2026, \bibinfo{title}{First Observation of a Polar Coronal Hole-like Fast Solar Wind Stream in the Sub-Alfv\'enic Solar Corona: An Analysis of Turbulence Properties,} Under Review in the Astrophysical Journal Letters

\bibitem[{J.~W. {Belcher} \& J. {Davis}(1971){Belcher} \& {Davis}}]{Belcher1971}
{Belcher}, J.~W., \& {Davis}, Leverett, J. 1971, \bibinfo{title}{{Large-amplitude Alfv{\'e}n waves in the interplanetary medium, 2},} \jgr, 76, 3534, \dodoi{10.1029/JA076i016p03534}

\bibitem[{T.~A. {Bowen} {et~al.}(2025){Bowen}, {Mallet}, {Dunn}, {Squire}, {Chandran}, {Meyrand}, {Davis}, {Dudok de Wit}, {Bale}, {Badman}, \& {Sioulas}}]{Bowen2025}
{Bowen}, T.~A., {Mallet}, A., {Dunn}, C.~I., {et~al.} 2025, \bibinfo{title}{{Formation of magnetic switchbacks via expanding Alfv{\'e}n waves},} \aap, 700, A51, \dodoi{10.1051/0004-6361/202450220}

\bibitem[{A.~R. {Breen} {et~al.}(1996){Breen}, {Coles}, {Grall}, {Klinglesmith}, {Markkanen}, {Moran}, {Tegid}, \& {Williams}}]{Breen1996}
{Breen}, A.~R., {Coles}, W.~A., {Grall}, R.~R., {et~al.} 1996, \bibinfo{title}{{EISCAT measurements of the solar wind},} Annales Geophysicae, 14, 1235, \dodoi{10.1007/s00585-996-1235-8}

\bibitem[{A.~W. {Case} {et~al.}(2020){Case}, {Kasper}, {Stevens}, {Korreck}, {Paulson}, {Daigneau}, {Caldwell}, {Freeman}, {Henry}, {Klingensmith}, {Bookbinder}, {Robinson}, {Berg}, {Tiu}, {Wright}, {Reinhart}, {Curtis}, {Ludlam}, {Larson}, {Whittlesey}, {Livi}, {Klein}, \& {Martinovi{\'c}}}]{Case2020}
{Case}, A.~W., {Kasper}, J.~C., {Stevens}, M.~L., {et~al.} 2020, \bibinfo{title}{{The Solar Probe Cup on the Parker Solar Probe},} \apjs, 246, 43, \dodoi{10.3847/1538-4365/ab5a7b}

\bibitem[{B.~D.~G. {Chandran} {et~al.}(2025){Chandran}, {Adkins}, {Bale}, {David}, {Halekas}, {Klein}, {Meyrand}, {Perez}, {Shoda}, {Squire}, \& {Yerger}}]{Chandran2025}
{Chandran}, B. D.~G., {Adkins}, T., {Bale}, S.~D., {et~al.} 2025, \bibinfo{title}{{A two-fluid solar-wind model with intermittent Alfv{\'e}nic turbulence},} Journal of Plasma Physics, 91, E125, \dodoi{10.1017/S0022377825100640}

\bibitem[{S.~R. {Cranmer}(2020{\natexlab{a}}){Cranmer}}]{Cranmer2020a}
{Cranmer}, S.~R. 2020{\natexlab{a}}, \bibinfo{title}{{Heating Rates for Protons and Electrons in Polar Coronal Holes: Empirical Constraints from the Ultraviolet Coronagraph Spectrometer},} \apj, 900, 105, \dodoi{10.3847/1538-4357/abab04}

\bibitem[{S.~R. {Cranmer}(2020{\natexlab{b}}){Cranmer}}]{Cranmer2020b}
{Cranmer}, S.~R. 2020{\natexlab{b}}, \bibinfo{title}{{Updated Measurements of Proton, Electron, and Oxygen Temperatures in the Fast Solar Wind},} Research Notes of the American Astronomical Society, 4, 249, \dodoi{10.3847/2515-5172/abd5ae}

\bibitem[{S.~R. {Cranmer} \& A.~A. {van Ballegooijen}(2005){Cranmer} \& {van Ballegooijen}}]{Cranmer2005}
{Cranmer}, S.~R., \& {van Ballegooijen}, A.~A. 2005, \bibinfo{title}{{On the Generation, Propagation, and Reflection of Alfv{\'e}n Waves from the Solar Photosphere to the Distant Heliosphere},} \apjs, 156, 265, \dodoi{10.1086/426507}

\bibitem[{S.~R. {Cranmer} {et~al.}(2007){Cranmer}, {van Ballegooijen}, \& {Edgar}}]{Cranmer2007}
{Cranmer}, S.~R., {van Ballegooijen}, A.~A., \& {Edgar}, R.~J. 2007, \bibinfo{title}{{Self-consistent Coronal Heating and Solar Wind Acceleration from Anisotropic Magnetohydrodynamic Turbulence},} \apjs, 171, 520, \dodoi{10.1086/518001}

\bibitem[{J.-B. {Dakeyo} {et~al.}(2025){Dakeyo}, {D{\'e}moulin}, {Rouillard}, {Maksimovic}, {Chapiron}, \& {Bale}}]{Dakeyo2025_bipoly}
{Dakeyo}, J.-B., {D{\'e}moulin}, P., {Rouillard}, A., {et~al.} 2025, \bibinfo{title}{{Generalized Two Thermal Regime Approach: Bipoly Fluid Modeling},} \apj, 986, 157, \dodoi{10.3847/1538-4357/add473}

\bibitem[{J.-B. {Dakeyo} {et~al.}(2022){Dakeyo}, {Maksimovic}, {D{\'e}moulin}, {Halekas}, \& {Stevens}}]{Dakeyo2022}
{Dakeyo}, J.-B., {Maksimovic}, M., {D{\'e}moulin}, P., {Halekas}, J., \& {Stevens}, M.~L. 2022, \bibinfo{title}{{Statistical Analysis of the Radial Evolution of the Solar Winds between 0.1 and 1 au and Their Semiempirical Isopoly Fluid Modeling},} \apj, 940, 130, \dodoi{10.3847/1538-4357/ac9b14}

\bibitem[{G. {de Toma}(2011){de Toma}}]{DeToma2011}
{de Toma}, G. 2011, \bibinfo{title}{{Evolution of Coronal Holes and Implications for High-Speed Solar Wind During the Minimum Between Cycles 23 and 24},} \solphys, 274, 195, \dodoi{10.1007/s11207-010-9677-2}

\bibitem[{C.~E. {DeForest} {et~al.}(2018){DeForest}, {Howard}, {Velli}, {Viall}, \& {Vourlidas}}]{DeForest2018}
{DeForest}, C.~E., {Howard}, R.~A., {Velli}, M., {Viall}, N., \& {Vourlidas}, A. 2018, \bibinfo{title}{{The Highly Structured Outer Solar Corona},} \apj, 862, 18, \dodoi{10.3847/1538-4357/aac8e3}

\bibitem[{C.~E. {DeForest} {et~al.}(2026){DeForest}, {Gibson}, {Killough}, {Waltham}, {Beasley}, {Colaninno}, {Laurent}, {Seaton}, {Hughes}, {Guhathakurta}, {Viall}, {Atti{\'e}}, {Banerjee}, {Barnard}, {Biesecker}, {Bisi}, {Bothmer}, {Brody}, {Burkepile}, {Cairns}, {Campbell}, {Case}, {Caspi}, {Cheney}, {Chhiber}, {Clapp}, {Cranmer}, {Davies}, {de Koning}, {Desai}, {Elliott}, {Farid}, {Gallardo-Lacourt}, {Gilly}, {Gobat}, {Hanson}, {Harrison}, {Hassler}, {Henley}, {Henry}, {Howard}, {Jackson}, {Jones}, {Kolinski}, {Lamb}, {Lehtinen}, {Lowder}, {Malanushenko}, {Matthaeus}, {McComas}, {McGee}, {Morgan}, {Oberoi}, {Odstrcil}, {Parmenter}, {Patel}, {Pecora}, {Persyn}, {Pizzo}, {Plunkett}, {Provornikova}, {Raouafi}, {Redfern}, {Rouillard}, {Smith}, {Smith}, {Talpas}, {Tappin}, {Thernisien}, {Thompson}, {Van Kooten}, {Walsh}, {Webb}, {Wells}, {West}, {Wiens}, {Yang}, \& {Zhukov}}]{Deforest2026}
{DeForest}, C.~E., {Gibson}, S.~E., {Killough}, R., {et~al.} 2026, \bibinfo{title}{{Polarimeter to Unify the Corona and Heliosphere (PUNCH)},} \solphys, 301, 16, \dodoi{10.1007/s11207-026-02608-2}

\bibitem[{Y. {Ding} {et~al.}(2026){Ding}, {Velli}, {Huang}, {Shi}, {Matteini}, {Sioulas}, {Bale}, {Tenerani}, \& {Liu}}]{Ding26}
{Ding}, Y., {Velli}, M., {Huang}, Z., {et~al.} 2026, \bibinfo{title}{{Spherically Polarized Alfv{\'e}n Waves and the Gosling Boost},} arXiv e-prints, arXiv:2608.05091, \dodoi{10.48550/arXiv.2608.05091}

\bibitem[{J.~G. {Doyle} {et~al.}(2010){Doyle}, {Chapman}, {Bryans}, {P{\'e}rez-Su{\'a}rez}, {Singh}, {Summers}, \& {Savin}}]{Doyle2010}
{Doyle}, J.~G., {Chapman}, S., {Bryans}, P., {et~al.} 2010, \bibinfo{title}{{Deriving the coronal hole electron temperature: electron density dependent ionization / recombination considerations},} Research in Astronomy and Astrophysics, 10, 91, \dodoi{10.1088/1674-4527/10/1/008}

\bibitem[{N. {Fargette} {et~al.}(2021){Fargette}, {Lavraud}, {Rouillard}, {R{\'e}ville}, {Dudok De Wit}, {Froment}, {Halekas}, {Phan}, {Malaspina}, {Bale}, {Kasper}, {Louarn}, {Case}, {Korreck}, {Larson}, {Pulupa}, {Stevens}, {Whittlesey}, \& {Berthomier}}]{Fargette2021}
{Fargette}, N., {Lavraud}, B., {Rouillard}, A.~P., {et~al.} 2021, \bibinfo{title}{{Characteristic Scales of Magnetic Switchback Patches Near the Sun and Their Possible Association With Solar Supergranulation and Granulation},} \apj, 919, 96, \dodoi{10.3847/1538-4357/ac1112}

\bibitem[{A.~J. {Finley}(2025){Finley}}]{Finley2025}
{Finley}, A.~J. 2025, \bibinfo{title}{{Reconstructing the Sun's Alfv{\'e}n surface and wind braking torque with Parker Solar Probe},} \aap, 702, A252, \dodoi{10.1051/0004-6361/202556391}

\bibitem[{A.~J. Finley {et~al.}(2026)Finley, Brun, Strugarek, Perri, Müller, Janvier, To, \& Eklund}]{finley2026arxiv}
Finley, A.~J., Brun, A.~S., Strugarek, A., {et~al.} 2026, \bibinfo{title}{Active nests at solar maximum: How nested flux emergence dominated flaring activity and structured the heliosphere,} \doarXiv{2608.26315}

\bibitem[{A.~J. {Finley} {et~al.}(2019){Finley}, {Hewitt}, {Matt}, {Owens}, {Pinto}, \& {R{\'e}ville}}]{FInley2019}
{Finley}, A.~J., {Hewitt}, A.~L., {Matt}, S.~P., {et~al.} 2019, \bibinfo{title}{{Direct Detection of Solar Angular Momentum Loss with the Wind Spacecraft},} \apjl, 885, L30, \dodoi{10.3847/2041-8213/ab4ff4}

\bibitem[{N.~J. {Fox} {et~al.}(2016){Fox}, {Velli}, {Bale}, {Decker}, {Driesman}, {Howard}, {Kasper}, {Kinnison}, {Kusterer}, {Lario}, {Lockwood}, {McComas}, {Raouafi}, \& {Szabo}}]{Fox2016}
{Fox}, N.~J., {Velli}, M.~C., {Bale}, S.~D., {et~al.} 2016, \bibinfo{title}{{The Solar Probe Plus Mission: Humanity's First Visit to Our Star},} \ssr, 204, 7, \dodoi{10.1007/s11214-015-0211-6}

\bibitem[{T.~M. {Garton} {et~al.}(2018){Garton}, {Murray}, \& {Gallagher}}]{Garton2018}
{Garton}, T.~M., {Murray}, S.~A., \& {Gallagher}, P.~T. 2018, \bibinfo{title}{{Expansion of High-speed Solar Wind Streams from Coronal Holes through the Inner Heliosphere},} \apjl, 869, L12, \dodoi{10.3847/2041-8213/aaf39a}

\bibitem[{V. G{\'e}not \& S.~J. Schwartz(2004)G{\'e}not \& Schwartz}]{genot2004}
G{\'e}not, V., \& Schwartz, S.~J. 2004, \bibinfo{title}{Spacecraft potential effects on electron moments derived from a perfect plasma detector,} Annales Geophysicae, 22, 2073, \dodoi{10.5194/angeo-22-2073-2004}

\bibitem[{S. {Giordano} {et~al.}(2025){Giordano}, {Spadaro}, {Susino}, {Ventura}, {Zangrilli}, {Andretta}, {De Leo}, {Romoli}, {Teriaca}, {Uslenghi}, {Fineschi}, {Telloni}, {Landini}, {Nicolini}, {Pancrazzi}, \& {Sasso}}]{Giordano2025}
{Giordano}, S., {Spadaro}, D., {Susino}, R., {et~al.} 2025, \bibinfo{title}{{Solar wind speed maps from the Metis coronagraph observations},} \aap, 701, A56, \dodoi{10.1051/0004-6361/202554105}

\bibitem[{T.~I. {Gombosi} {et~al.}(2018){Gombosi}, {van der Holst}, {Manchester}, \& {Sokolov}}]{Gombosi2018}
{Gombosi}, T.~I., {van der Holst}, B., {Manchester}, W.~B., \& {Sokolov}, I.~V. 2018, \bibinfo{title}{{Extended MHD modeling of the steady solar corona and the solar wind},} Living Reviews in Solar Physics, 15, 4, \dodoi{10.1007/s41116-018-0014-4}

\bibitem[{J. {Goodwill} {et~al.}(2026){Goodwill}, {Adhikari}, {Payne}, {Bandyopadhyay}, {Badman}, {Pecora}, {Pongkitiwanichakul}, {Pradata}, {Romeo}, {Roy}, {Ruffolo}, {Stevens}, {Thepthong}, {Usmanov}, {Wang}, {Goldstein}, {Chhiber}, \& {Matthaeus}}]{Goodwill2026}
{Goodwill}, J., {Adhikari}, S., {Payne}, D., {et~al.} 2026, \bibinfo{title}{{Parker Solar Probe analysis across the Alfv{\'e}nic transition: velocity shear, magnetic deflection, and switchback formation},} \mnras, 547, stag242, \dodoi{10.1093/mnras/stag242}

\bibitem[{J.~T. {Gosling} {et~al.}(2009){Gosling}, {McComas}, {Roberts}, \& {Skoug}}]{Gosling2009}
{Gosling}, J.~T., {McComas}, D.~J., {Roberts}, D.~A., \& {Skoug}, R.~M. 2009, \bibinfo{title}{{A One-Sided Aspect of Alfvenic Fluctuations in the Solar Wind},} \apjl, 695, L213, \dodoi{10.1088/0004-637X/695/2/L213}

\bibitem[{R.~R. {Grall} {et~al.}(1996){Grall}, {Coles}, {Klinglesmith}, {Breen}, {Williams}, {Markkanen}, \& {Esser}}]{Grall1996}
{Grall}, R.~R., {Coles}, W.~A., {Klinglesmith}, M.~T., {et~al.} 1996, \bibinfo{title}{{Rapid acceleration of the polar solar wind},} \nat, 379, 429, \dodoi{10.1038/379429a0}

\bibitem[{Y. {Guo} {et~al.}(2021){Guo}, {Thompson}, {Wirzburger}, {Pinkine}, {Bushman}, {Goodson}, {Haw}, {Hudson}, {Jones}, {Kijewski}, {Lathrop}, {Lau}, {Mottinger}, {Ryne}, {Shyong}, {Valerino}, \& {Whittenburg}}]{Guo2021}
{Guo}, Y., {Thompson}, P., {Wirzburger}, J., {et~al.} 2021, \bibinfo{title}{{Execution of Parker Solar Probe's unprecedented flight to the Sun and early results},} Acta Astronautica, 179, 425, \dodoi{10.1016/j.actaastro.2020.11.007}

\bibitem[{J.~S. {Halekas} {et~al.}(2020){Halekas}, {Whittlesey}, {Larson}, {McGinnis}, {Maksimovic}, {Berthomier}, {Kasper}, {Case}, {Korreck}, {Stevens}, {Klein}, {Bale}, {MacDowall}, {Pulupa}, {Malaspina}, {Goetz}, \& {Harvey}}]{Halekas2020}
{Halekas}, J.~S., {Whittlesey}, P., {Larson}, D.~E., {et~al.} 2020, \bibinfo{title}{{Electrons in the Young Solar Wind: First Results from the Parker Solar Probe},} \apjs, 246, 22, \dodoi{10.3847/1538-4365/ab4cec}

\bibitem[{J.~S. {Halekas} {et~al.}(2022){Halekas}, {Whittlesey}, {Larson}, {Maksimovic}, {Livi}, {Berthomier}, {Kasper}, {Case}, {Stevens}, {Bale}, {MacDowall}, \& {Pulupa}}]{Halekas2022}
{Halekas}, J.~S., {Whittlesey}, P., {Larson}, D.~E., {et~al.} 2022, \bibinfo{title}{{The Radial Evolution of the Solar Wind as Organized by Electron Distribution Parameters},} \apj, 936, 53, \dodoi{10.3847/1538-4357/ac85b8}

\bibitem[{J.~S. {Halekas} {et~al.}(2023){Halekas}, {Bale}, {Berthomier}, {Chandran}, {Drake}, {Kasper}, {Klein}, {Larson}, {Livi}, {Pulupa}, {Stevens}, {Verniero}, \& {Whittlesey}}]{Halekas2023}
{Halekas}, J.~S., {Bale}, S.~D., {Berthomier}, M., {et~al.} 2023, \bibinfo{title}{{Quantifying the Energy Budget in the Solar Wind from 13.3 to 100 Solar Radii},} \apj, 952, 26, \dodoi{10.3847/1538-4357/acd769}

\bibitem[{J.~K. {Harmon} \& W.~A. {Coles}(2005){Harmon} \& {Coles}}]{Harmon2005}
{Harmon}, J.~K., \& {Coles}, W.~A. 2005, \bibinfo{title}{{Modeling radio scattering and scintillation observations of the inner solar wind using oblique Alfv{\'e}n/ion cyclotron waves},} Journal of Geophysical Research (Space Physics), 110, A03101, \dodoi{10.1029/2004JA010834}

\bibitem[{M. {Hegde} \& K.~M. {Hiremath}(2024){Hegde} \& {Hiremath}}]{Hegde2024}
{Hegde}, M., \& {Hiremath}, K.~M. 2024, \bibinfo{title}{{Thermal and magnetic field structure of near-equatorial coronal holes},} \aap, 688, A35, \dodoi{10.1051/0004-6361/202347082}

\bibitem[{K.~S. {Hickmann} {et~al.}(2015){Hickmann}, {Godinez}, {Henney}, \& {Arge}}]{Hickmann2015}
{Hickmann}, K.~S., {Godinez}, H.~C., {Henney}, C.~J., \& {Arge}, C.~N. 2015, \bibinfo{title}{{Data Assimilation in the ADAPT Photospheric Flux Transport Model},} \solphys, 290, 1105, \dodoi{10.1007/s11207-015-0666-3}

\bibitem[{S.~J. {Hofmeister}(2026){Hofmeister}}]{Hofmeister2026}
{Hofmeister}, S.~J. 2026, \bibinfo{title}{{MHD simulations on the large-scale propagation of high-speed solar wind streams},} arXiv e-prints, arXiv:2605.01613, \dodoi{10.48550/arXiv.2605.01613}

\bibitem[{S.~J. {Hofmeister} {et~al.}(2018){Hofmeister}, {Veronig}, {Temmer}, {Vennerstrom}, {Heber}, \& {Vr{\v{s}}nak}}]{Hofmeister2018}
{Hofmeister}, S.~J., {Veronig}, A., {Temmer}, M., {et~al.} 2018, \bibinfo{title}{{The Dependence of the Peak Velocity of High-Speed Solar Wind Streams as Measured in the Ecliptic by ACE and the STEREO satellites on the Area and Co-latitude of Their Solar Source Coronal Holes},} Journal of Geophysical Research (Space Physics), 123, 1738, \dodoi{10.1002/2017JA024586}

\bibitem[{S.~J. {Hofmeister} {et~al.}(2022){Hofmeister}, {Asvestari}, {Guo}, {Heidrich-Meisner}, {Heinemann}, {Magdalenic}, {Poedts}, {Samara}, {Temmer}, {Vennerstrom}, {Veronig}, {Vr{\v{s}}nak}, \& {Wimmer-Schweingruber}}]{Hofmeister2022}
{Hofmeister}, S.~J., {Asvestari}, E., {Guo}, J., {et~al.} 2022, \bibinfo{title}{{How the area of solar coronal holes affects the properties of high-speed solar wind streams near Earth: An analytical model},} \aap, 659, A190, \dodoi{10.1051/0004-6361/202141919}

\bibitem[{T.~S. {Horbury} {et~al.}(2020){Horbury}, {O'Brien}, {Carrasco Blazquez}, {Bendyk}, {Brown}, {Hudson}, {Evans}, {Oddy}, {Carr}, {Beek}, {Cupido}, {Bhattacharya}, {Dominguez}, {Matthews}, {Myklebust}, {Whiteside}, {Bale}, {Baumjohann}, {Burgess}, {Carbone}, {Cargill}, {Eastwood}, {Erd{\"o}s}, {Fletcher}, {Forsyth}, {Giacalone}, {Glassmeier}, {Goldstein}, {Hoeksema}, {Lockwood}, {Magnes}, {Maksimovic}, {Marsch}, {Matthaeus}, {Murphy}, {Nakariakov}, {Owen}, {Owens}, {Rodriguez-Pacheco}, {Richter}, {Riley}, {Russell}, {Schwartz}, {Vainio}, {Velli}, {Vennerstrom}, {Walsh}, {Wimmer-Schweingruber}, {Zank}, {M{\"u}ller}, {Zouganelis}, \& {Walsh}}]{Horbury2020}
{Horbury}, T.~S., {O'Brien}, H., {Carrasco Blazquez}, I., {et~al.} 2020, \bibinfo{title}{{The Solar Orbiter magnetometer},} \aap, 642, A9, \dodoi{10.1051/0004-6361/201937257}

\bibitem[{T.~S. {Horbury} {et~al.}(2023){Horbury}, {Bale}, {McManus}, {Larson}, {Kasper}, {Laker}, {Matteini}, {Raouafi}, {Velli}, {Woodham}, {Woolley}, {Fedorov}, {Louarn}, {Kieokaew}, {Durovcova}, {Chandran}, \& {Owen}}]{Horbury2023}
{Horbury}, T.~S., {Bale}, S.~D., {McManus}, M.~D., {et~al.} 2023, \bibinfo{title}{{Switchbacks, microstreams, and broadband turbulence in the solar wind},} Physics of Plasmas, 30, 082905, \dodoi{10.1063/5.0123250}

\bibitem[{J. {Huang} {et~al.}(2025){Huang}, {Larson}, {Ervin}, {Liu}, {Ortiz}, {Martinovi{\'c}}, {Huang}, {Chasapis}, {Chu}, {Alterman}, {Huang}, {Wei}, {Verniero}, {Jian}, {Szabo}, {Romeo}, {Rahmati}, {Livi}, {Whittlesey}, {Alnussirat}, {Kasper}, {Stevens}, \& {Bale}}]{Huang-2025}
{Huang}, J., {Larson}, D.~E., {Ervin}, T., {et~al.} 2025, \bibinfo{title}{{The Temperature Anisotropy and Helium Abundance Features of Alfv{\'e}nic Slow Solar Wind Observed by Parker Solar Probe, Helios, and Wind Missions},} \apjl, 986, L28, \dodoi{10.3847/2041-8213/ade0ac}

\bibitem[{K. {Jaffarove} {et~al.}(2026){Jaffarove}, {Ervin}, {Bale}, {Velli}, {Badman}, {Larson}, {Livi}, {Rivera}, \& {Romeo}}]{Jaffarove2026}
{Jaffarove}, K., {Ervin}, T., {Bale}, S.~D., {et~al.} 2026, \bibinfo{title}{{Connecting In-Situ Quiescent Regions to Solar Sources of a Sub-Alfv{\'e}nic Fast Wind Stream using Parker Solar Probe Observations},} arXiv e-prints, arXiv:2511.21971, \dodoi{10.48550/arXiv.2511.21971}

\bibitem[{M. {Johnson} {et~al.}(2024){Johnson}, {Rivera}, {Niembro}, {Paulson}, {Badman}, {Stevens}, {Dieguez}, {Case}, {Bale}, \& {Kasper}}]{Johnson2024}
{Johnson}, M., {Rivera}, Y.~J., {Niembro}, T., {et~al.} 2024, \bibinfo{title}{{Helium Abundance Periods Observed by the Solar Probe Cup on Parker Solar Probe: Encounters 1{\textendash}14},} \apj, 964, 81, \dodoi{10.3847/1538-4357/ad2510}

\bibitem[{M.~L. {Kaiser} {et~al.}(2008){Kaiser}, {Kucera}, {Davila}, {St. Cyr}, {Guhathakurta}, \& {Christian}}]{Kaiser2008}
{Kaiser}, M.~L., {Kucera}, T.~A., {Davila}, J.~M., {et~al.} 2008, \bibinfo{title}{{The STEREO Mission: An Introduction},} \ssr, 136, 5, \dodoi{10.1007/s11214-007-9277-0}

\bibitem[{J.~C. {Kasper} {et~al.}(2019){Kasper}, {Bale}, {Belcher}, {Berthomier}, {Case}, {Chandran}, {Curtis}, {Gallagher}, {Gary}, {Golub}, {Halekas}, {Ho}, {Horbury}, {Hu}, {Huang}, {Klein}, {Korreck}, {Larson}, {Livi}, {Maruca}, {Lavraud}, {Louarn}, {Maksimovic}, {Martinovic}, {McGinnis}, {Pogorelov}, {Richardson}, {Skoug}, {Steinberg}, {Stevens}, {Szabo}, {Velli}, {Whittlesey}, {Wright}, {Zank}, {MacDowall}, {McComas}, {McNutt}, {Pulupa}, {Raouafi}, \& {Schwadron}}]{Kasper2019}
{Kasper}, J.~C., {Bale}, S.~D., {Belcher}, J.~W., {et~al.} 2019, \bibinfo{title}{{Alfv{\'e}nic velocity spikes and rotational flows in the near-Sun solar wind},} \nat, 576, 228, \dodoi{10.1038/s41586-019-1813-z}

\bibitem[{J.~L. {Kohl} {et~al.}(1995){Kohl}, {Esser}, {Gardner}, {Habbal}, {Daigneau}, {Dennis}, {Nystrom}, {Panasyuk}, {Raymond}, {Smith}, {Strachan}, {Van Ballegooijen}, {Noci}, {Fineschi}, {Romoli}, {Ciaravella}, {Modigliani}, {Huber}, {Antonucci}, {Benna}, {Giordano}, {Tondello}, {Nicolosi}, {Naletto}, {Pernechele}, {Spadaro}, {Poletto}, {Livi}, {Von Der L{\"u}he}, {Geiss}, {Timothy}, {Gloeckler}, {Allegra}, {Basile}, {Brusa}, {Wood}, {Siegmund}, {Fowler}, {Fisher}, \& {Jhabvala}}]{Kohl1995}
{Kohl}, J.~L., {Esser}, R., {Gardner}, L.~D., {et~al.} 1995, \bibinfo{title}{{The Ultraviolet Coronagraph Spectrometer for the Solar and Heliospheric Observatory},} \solphys, 162, 313, \dodoi{10.1007/BF00733433}

\bibitem[{A. {Koval} {et~al.}(2021){Koval}, {Lepping}, \& {Szabo}}]{windmfi3s21a}
{Koval}, A., {Lepping}, R.~P., \& {Szabo}, A. 2021, \bibinfo{title}{\emph{Wind} Magnetic Field Investigation (MFI) Composite Data,}, 2.3.2 NASA Space Physics Data Facility, \dodoi{10.48322/av38-wn55}

\bibitem[{E. {Leer} \& T.~E. {Holzer}(1980){Leer} \& {Holzer}}]{Leer-Holzer-1980-sonicpoint}
{Leer}, E., \& {Holzer}, T.~E. 1980, \bibinfo{title}{{Energy addition in the solar wind.},} \jgr, 85, 4681, \dodoi{10.1029/JA085iA09p04681}

\bibitem[{E. {Leer} {et~al.}(1982){Leer}, {Holzer}, \& {Fla}}]{Leer1982}
{Leer}, E., {Holzer}, T.~E., \& {Fla}, T. 1982, \bibinfo{title}{{Acceleration of the solar wind.},} \ssr, 33, 161, \dodoi{10.1007/BF00213253}

\bibitem[{J.~R. {Lemen} {et~al.}(2012){Lemen}, {Title}, {Akin}, {Boerner}, {Chou}, {Drake}, {Duncan}, {Edwards}, {Friedlaender}, {Heyman}, {Hurlburt}, {Katz}, {Kushner}, {Levay}, {Lindgren}, {Mathur}, {McFeaters}, {Mitchell}, {Rehse}, {Schrijver}, {Springer}, {Stern}, {Tarbell}, {Wuelser}, {Wolfson}, {Yanari}, {Bookbinder}, {Cheimets}, {Caldwell}, {Deluca}, {Gates}, {Golub}, {Park}, {Podgorski}, {Bush}, {Scherrer}, {Gummin}, {Smith}, {Auker}, {Jerram}, {Pool}, {Soufli}, {Windt}, {Beardsley}, {Clapp}, {Lang}, \& {Waltham}}]{Lemen2012}
{Lemen}, J.~R., {Title}, A.~M., {Akin}, D.~J., {et~al.} 2012, \bibinfo{title}{{The Atmospheric Imaging Assembly (AIA) on the Solar Dynamics Observatory (SDO)},} \solphys, 275, 17, \dodoi{10.1007/s11207-011-9776-8}

\bibitem[{R.~P. {Lepping} {et~al.}(1995){Lepping}, {Ac{\~u}na}, {Burlaga}, {Farrell}, {Slavin}, {Schatten}, {Mariani}, {Ness}, {Neubauer}, {Whang}, {Byrnes}, {Kennon}, {Panetta}, {Scheifele}, \& {Worley}}]{lepping95}
{Lepping}, R.~P., {Ac{\~u}na}, M.~H., {Burlaga}, L.~F., {et~al.} 1995, \bibinfo{title}{{The Wind Magnetic Field Investigation},} Space Sci. Rev., 71, 207, \dodoi{10.1007/BF00751330}

\bibitem[{R.~P. {Lin} {et~al.}(1995){Lin}, {Anderson}, {Ashford}, {Carlson}, {Curtis}, {Ergun}, {Larson}, {McFadden}, {McCarthy}, {Parks}, {R\`{e}me}, {Bosqued}, {Coutelier}, {Cotin}, {D'Uston}, {Wenzel}, {Sanderson}, {Henrion}, {Ronnet}, \& {Paschmann}}]{lin95a}
{Lin}, R.~P., {Anderson}, K.~A., {Ashford}, S., {et~al.} 1995, \bibinfo{title}{{A Three-Dimensional Plasma and Energetic Particle Investigation for the Wind Spacecraft},} Space Sci. Rev., 71, 125, \dodoi{10.1007/BF00751328}

\bibitem[{R. {Lionello} {et~al.}(2014){Lionello}, {Velli}, {Downs}, {Linker}, {Miki{\'c}}, \& {Verdini}}]{Lionello2014}
{Lionello}, R., {Velli}, M., {Downs}, C., {et~al.} 2014, \bibinfo{title}{{Validating a Time-dependent Turbulence-driven Model of the Solar Wind},} \apj, 784, 120, \dodoi{10.1088/0004-637X/784/2/120}

\bibitem[{R. {Livi} {et~al.}(2022){Livi}, {Larson}, {Kasper}, {Abiad}, {Case}, {Klein}, {Curtis}, {Dalton}, {Stevens}, {Korreck}, {Ho}, {Robinson}, {Tiu}, {Whittlesey}, {Verniero}, {Halekas}, {McFadden}, {Marckwordt}, {Slagle}, {Abatcha}, {Rahmati}, \& {McManus}}]{Livi2022}
{Livi}, R., {Larson}, D.~E., {Kasper}, J.~C., {et~al.} 2022, \bibinfo{title}{{The Solar Probe ANalyzer-Ions on the Parker Solar Probe},} \apj, 938, 138, \dodoi{10.3847/1538-4357/ac93f5}

\bibitem[{P. {Louarn} {et~al.}(2021){Louarn}, {Fedorov}, {Prech}, {Owen}, {Bruno}, {Livi}, {Lavraud}, {Rouillard}, {G{\'e}not}, {Andr{\'e}}, {Fruit}, {R{\'e}ville}, {Kieokaew}, {Plotnikov}, {Penou}, {Barthe}, {Khataria}, {Berthomier}, {D'Amicis}, {Sorriso-Valvo}, {Allegrini}, {Raines}, {Verscharen}, {Fortunato}, {Mele}, {Horbury}, {O'brien}, {Evans}, {Angelini}, {Maksimovic}, {Kasper}, \& {Bale}}]{Louarn2021}
{Louarn}, P., {Fedorov}, A., {Prech}, L., {et~al.} 2021, \bibinfo{title}{{Multiscale views of an Alfv{\'e}nic slow solar wind: 3D velocity distribution functions observed by the Proton-Alpha Sensor of Solar Orbiter},} \aap, 656, A36, \dodoi{10.1051/0004-6361/202141095}

\bibitem[{A.~R. {Macneil} {et~al.}(2020){Macneil}, {Owens}, {Wicks}, {Lockwood}, {Bentley}, \& {Lang}}]{Macneil2020}
{Macneil}, A.~R., {Owens}, M.~J., {Wicks}, R.~T., {et~al.} 2020, \bibinfo{title}{{The evolution of inverted magnetic fields through the inner heliosphere},} \mnras, 494, 3642, \dodoi{10.1093/mnras/staa951}

\bibitem[{M. Maksimovic {et~al.}(2020)Maksimovic, Bale, Chust, Khotyaintsev, Krasnoselskikh, Kretzschmar, Plettemeier, Rucker, Sou{\v{c}}ek, Steller, {et~al.}}]{maksimovic_rpw}
Maksimovic, M., Bale, S.~D., Chust, T., {et~al.} 2020, \bibinfo{title}{The {Solar Orbiter Radio and Plasma Waves (RPW)} instrument,} Astronomy \& Astrophysics, 642, A12, \dodoi{10.1051/0004-6361/201936214}

\bibitem[{A. {Mallet} {et~al.}(2026){Mallet}, {Shi}, {Tenerani}, {Agapitov}, {Akhavan-Tafti}, {Badman}, {Bizien}, {Bowen}, {Desai}, {Drake}, {Horbury}, {Larosa}, {Madjarska}, {Malara}, {Matteini}, {Owens}, {R{\'e}ville}, {Sioulas}, {Soni}, {Squire}, {Suen}, {Swisdak}, {Velli}, {Verniero}, {Watkins}, \& {Sorriso-Valvo}}]{Mallet2026}
{Mallet}, A., {Shi}, C., {Tenerani}, A., {et~al.} 2026, \bibinfo{title}{{Evolution and Impact of Switchbacks Throughout the Heliosphere},} \ssr, 222, 59, \dodoi{10.1007/s11214-026-01311-9}

\bibitem[{M.~M. {Martinovi{\'c}} {et~al.}(2020){Martinovi{\'c}}, {Klein}, {Gramze}, {Jain}, {Maksimovi{\'c}}, {Zaslavsky}, {Salem}, {Zouganelis}, \& {Simi{\'c}}}]{martinovic20a}
{Martinovi{\'c}}, M.~M., {Klein}, K.~G., {Gramze}, S.~R., {et~al.} 2020, \bibinfo{title}{{Solar Wind Electron Parameters Determination on Wind Spacecraft Using Quasi-Thermal Noise Spectroscopy},} J. Geophys. Res., 125, e28113, \dodoi{10.1029/2020JA028113}

\bibitem[{L. {Matteini} {et~al.}(2014){Matteini}, {Horbury}, {Neugebauer}, \& {Goldstein}}]{Matteini2014}
{Matteini}, L., {Horbury}, T.~S., {Neugebauer}, M., \& {Goldstein}, B.~E. 2014, \bibinfo{title}{{Dependence of solar wind speed on the local magnetic field orientation: Role of Alfv{\'e}nic fluctuations},} \grl, 41, 259, \dodoi{10.1002/2013GL058482}

\bibitem[{L. {Matteini} {et~al.}(2015){Matteini}, {Horbury}, {Pantellini}, {Velli}, \& {Schwartz}}]{Matteini2015}
{Matteini}, L., {Horbury}, T.~S., {Pantellini}, F., {Velli}, M., \& {Schwartz}, S.~J. 2015, \bibinfo{title}{{Ion Kinetic Energy Conservation and Magnetic Field Strength Constancy in Multi-fluid Solar Wind Alfv{\'e}nic Turbulence},} \apj, 802, 11, \dodoi{10.1088/0004-637X/802/1/11}

\bibitem[{W.~H. {Matthaeus} {et~al.}(1999){Matthaeus}, {Zank}, {Oughton}, {Mullan}, \& {Dmitruk}}]{Matthaeus1999}
{Matthaeus}, W.~H., {Zank}, G.~P., {Oughton}, S., {Mullan}, D.~J., \& {Dmitruk}, P. 1999, \bibinfo{title}{{Coronal Heating by Magnetohydrodynamic Turbulence Driven by Reflected Low-Frequency Waves},} \apjl, 523, L93, \dodoi{10.1086/312259}

\bibitem[{D.~J. {McComas} {et~al.}(2000){McComas}, {Barraclough}, {Funsten}, {Gosling}, {Santiago-Mu{\~n}oz}, {Skoug}, {Goldstein}, {Neugebauer}, {Riley}, \& {Balogh}}]{McComas2000}
{McComas}, D.~J., {Barraclough}, B.~L., {Funsten}, H.~O., {et~al.} 2000, \bibinfo{title}{{Solar wind observations over Ulysses' first full polar orbit},} \jgr, 105, 10419, \dodoi{10.1029/1999JA000383}

\bibitem[{M.~D. {McManus} {et~al.}(2022){McManus}, {Verniero}, {Bale}, {Bowen}, {Larson}, {Kasper}, {Livi}, {Matteini}, {Rahmati}, {Romeo}, {Whittlesey}, \& {Woolley}}]{McManus2022}
{McManus}, M.~D., {Verniero}, J., {Bale}, S.~D., {et~al.} 2022, \bibinfo{title}{{Density and Velocity Fluctuations of Alpha Particles in Magnetic Switchbacks},} \apj, 933, 43, \dodoi{10.3847/1538-4357/ac6ba3}

\bibitem[{M.~P. {Miralles} {et~al.}(2004){Miralles}, {Cranmer}, \& {Kohl}}]{Miralles2004}
{Miralles}, M.~P., {Cranmer}, S.~R., \& {Kohl}, J.~L. 2004, \bibinfo{title}{{Low-latitude coronal holes during solar maximum},} Advances in Space Research, 33, 696, \dodoi{10.1016/S0273-1177(03)00239-4}

\bibitem[{M.~P. {Miralles} {et~al.}(2001){Miralles}, {Cranmer}, {Panasyuk}, {Romoli}, \& {Kohl}}]{Miralles2001}
{Miralles}, M.~P., {Cranmer}, S.~R., {Panasyuk}, A.~V., {Romoli}, M., \& {Kohl}, J.~L. 2001, \bibinfo{title}{{Comparison of Empirical Models for Polar and Equatorial Coronal Holes},} \apjl, 549, L257, \dodoi{10.1086/319166}

\bibitem[{M. {Moncuquet} {et~al.}(2020){Moncuquet}, {Meyer-Vernet}, {Issautier}, {Pulupa}, {Bonnell}, {Bale}, {Dudok de Wit}, {Goetz}, {Griton}, {Harvey}, {MacDowall}, {Maksimovic}, \& {Malaspina}}]{Moncuquet2020}
{Moncuquet}, M., {Meyer-Vernet}, N., {Issautier}, K., {et~al.} 2020, \bibinfo{title}{{First In Situ Measurements of Electron Density and Temperature from Quasi-thermal Noise Spectroscopy with Parker Solar Probe/FIELDS},} \apjs, 246, 44, \dodoi{10.3847/1538-4365/ab5a84}

\bibitem[{P.~J. {Moran} {et~al.}(1997){Moran}, {Breen}, {Varley}, {Williams}, {Coles}, {Grall}, {Klinglesmith}, \& {Markkanen}}]{Moran1997}
{Moran}, P.~J., {Breen}, A.~R., {Varley}, C.~A., {et~al.} 1997, \bibinfo{title}{{EISCAT measurements of the solar wind: Measurements of the fast and slow streams},} Physics and Chemistry of the Earth, 22, 391, \dodoi{10.1016/S0079-1946(97)00164-X}

\bibitem[{M. {Neugebauer} {et~al.}(1995){Neugebauer}, {Goldstein}, {McComas}, {Suess}, \& {Balogh}}]{Neugebauer1995}
{Neugebauer}, M., {Goldstein}, B.~E., {McComas}, D.~J., {Suess}, S.~T., \& {Balogh}, A. 1995, \bibinfo{title}{{Ulysses Observations of Microstreams in the Solar Wind from Coronal Holes},} \jgr, 100, 23389, \dodoi{10.1029/95JA02723}

\bibitem[{G. {Nistic{\`o}} {et~al.}(2011){Nistic{\`o}}, {Patsourakos}, {Bothmer}, \& {Zimbardo}}]{Nistico2011}
{Nistic{\`o}}, G., {Patsourakos}, S., {Bothmer}, V., \& {Zimbardo}, G. 2011, \bibinfo{title}{{Determination of temperature maps of EUV coronal hole jets},} Advances in Space Research, 48, 1490, \dodoi{10.1016/j.asr.2011.07.003}

\bibitem[{J.~T. {Nolte} \& E.~C. {Roelof}(1973){Nolte} \& {Roelof}}]{Nolte1973}
{Nolte}, J.~T., \& {Roelof}, E.~C. 1973, \bibinfo{title}{{Large-Scale Structure of the Interplanetary Medium, I: High Coronal Source Longitude of the Quiet-Time Solar Wind},} \solphys, 33, 241, \dodoi{10.1007/BF00152395}

\bibitem[{J.~T. {Nolte} {et~al.}(1976){Nolte}, {Krieger}, {Timothy}, {Gold}, {Roelof}, {Vaiana}, {Lazarus}, {Sullivan}, \& {McIntosh}}]{Nolte1976}
{Nolte}, J.~T., {Krieger}, A.~S., {Timothy}, A.~F., {et~al.} 1976, \bibinfo{title}{{Coronal holes as sources of solar wind.},} \solphys, 46, 303, \dodoi{10.1007/BF00149859}

\bibitem[{K.~W. {Ogilvie} {et~al.}(2021){Ogilvie}, {Merka}, {Vinas}, \& {Fitzenreiter}}]{windswestrahlh521a}
{Ogilvie}, K.~W., {Merka}, J., {Vinas}, A.-F., \& {Fitzenreiter}, R.~J. 2021, \bibinfo{title}{\emph{Wind} Solar Wind Experiment (SWE) Electron Quadrature Moments Parameters (12--15s rate) (New Mode),}, 2.3.2 NASA Space Physics Data Facility, \dodoi{10.48322/chaz-z942}

\bibitem[{K.~W. {Ogilvie} {et~al.}(1995{\natexlab{a}}){Ogilvie}, {Chornay}, {Fritzenreiter}, {Hunsaker}, {Keller}, {Lobell}, {Miller}, {Scudder}, {Sittler}, {Torbert}, {Bodet}, {Needell}, {Lazarus}, {Steinberg}, {Tappan}, {Mavretic}, \& {Gergin}}]{Ogilvie1995}
{Ogilvie}, K.~W., {Chornay}, D.~J., {Fritzenreiter}, R.~J., {et~al.} 1995{\natexlab{a}}, \bibinfo{title}{{SWE, A Comprehensive Plasma Instrument for the Wind Spacecraft},} \ssr, 71, 55, \dodoi{10.1007/BF00751326}

\bibitem[{K.~W. {Ogilvie} {et~al.}(1995{\natexlab{b}}){Ogilvie}, {Chornay}, {Fritzenreiter}, {Hunsaker}, {Keller}, {Lobell}, {Miller}, {Scudder}, {Sittler}, {Torbert}, {Bodet}, {Needell}, {Lazarus}, {Steinberg}, {Tappan}, {Mavretic}, \& {Gergin}}]{ogilvie95}
{Ogilvie}, K.~W., {Chornay}, D.~J., {Fritzenreiter}, R.~J., {et~al.} 1995{\natexlab{b}}, \bibinfo{title}{{SWE, A Comprehensive Plasma Instrument for the Wind Spacecraft},} Space Sci. Rev., 71, 55, \dodoi{10.1007/BF00751326}

\bibitem[{C.~J. {Owen} {et~al.}(2020){Owen}, {Bruno}, {Livi}, {Louarn}, {Al Janabi}, {Allegrini}, {Amoros}, {Baruah}, {Barthe}, {Berthomier}, {Bordon}, {Brockley-Blatt}, {Brysbaert}, {Capuano}, {Collier}, {DeMarco}, {Fedorov}, {Ford}, {Fortunato}, {Fratter}, {Galvin}, {Hancock}, {Heirtzler}, {Kataria}, {Kistler}, {Lepri}, {Lewis}, {Loeffler}, {Marty}, {Mathon}, {Mayall}, {Mele}, {Ogasawara}, {Orlandi}, {Pacros}, {Penou}, {Persyn}, {Petiot}, {Phillips}, {P{\v{r}}ech}, {Raines}, {Reden}, {Rouillard}, {Rousseau}, {Rubiella}, {Seran}, {Spencer}, {Thomas}, {Trevino}, {Verscharen}, {Wurz}, {Alapide}, {Amoruso}, {Andr{\'e}}, {Anekallu}, {Arciuli}, {Arnett}, {Ascolese}, {Bancroft}, {Bland}, {Brysch}, {Calvanese}, {Castronuovo}, {{\v{C}}erm{\'a}k}, {Chornay}, {Clemens}, {Coker}, {Collinson}, {D'Amicis}, {Dandouras}, {Darnley}, {Davies}, {Davison}, {De Los Santos}, {Devoto}, {Dirks}, {Edlund}, {Fazakerley}, {Ferris}, {Frost}, {Fruit}, {Garat}, {G{\'e}not}, {Gibson}, {Gilbert}, {de Giosa}, {Gradone}, {Hailey},
  {Horbury}, {Hunt}, {Jacquey}, {Johnson}, {Lavraud}, {Lawrenson}, {Leblanc}, {Lockhart}, {Maksimovic}, {Malpus}, {Marcucci}, {Mazelle}, {Monti}, {Myers}, {Nguyen}, {Rodriguez-Pacheco}, {Phillips}, {Popecki}, {Rees}, {Rogacki}, {Ruane}, {Rust}, {Salatti}, {Sauvaud}, {Stakhiv}, {Stange}, {Stubbs}, {Taylor}, {Techer}, {Terrier}, {Thibodeaux}, {Urdiales}, {Varsani}, {Walsh}, {Watson}, {Wheeler}, {Willis}, {Wimmer-Schweingruber}, {Winter}, {Yardley}, \& {Zouganelis}}]{Owen2020}
{Owen}, C.~J., {Bruno}, R., {Livi}, S., {et~al.} 2020, \bibinfo{title}{{The Solar Orbiter Solar Wind Analyser (SWA) suite},} \aap, 642, A16, \dodoi{10.1051/0004-6361/201937259}

\bibitem[{D. {Payne}(et al. in prep.){Payne}}]{Payne2026_inprep}
{Payne}, D. et al. in prep., \bibinfo{title}{Large-Scale Thermodynamic Evolution of Coronal Hole Stream Interactions with Adjacent Slow Solar Wind,} in prep.

\bibitem[{D. {Payne} {et~al.}(2026){Payne}, {Akhavan-Tafti}, {Goodwill}, {Badman}, {Bandyopadhyay}, {Zank}, {Adhikari}, {Matthaeus}, {Shi}, {Stevens}, {Livi}, {Rivera}, \& {Paulson}}]{Payne2026}
{Payne}, D., {Akhavan-Tafti}, M., {Goodwill}, J., {et~al.} 2026, \bibinfo{title}{{Evolution of Magnetic Deflections at a Conversion Layer near the Alfv{\'e}n Surface},} \apjl, 1001, L29, \dodoi{10.3847/2041-8213/ae4fbd}

\bibitem[{F. {Pecora} {et~al.}(2022){Pecora}, {Matthaeus}, {Primavera}, {Greco}, {Chhiber}, {Bandyopadhyay}, \& {Servidio}}]{Pecora2022}
{Pecora}, F., {Matthaeus}, W.~H., {Primavera}, L., {et~al.} 2022, \bibinfo{title}{{Magnetic Switchback Occurrence Rates in the Inner Heliosphere: Parker Solar Probe and 1 au},} \apjl, 929, L10, \dodoi{10.3847/2041-8213/ac62d4}

\bibitem[{J.~L. {Phillips} {et~al.}(1995){Phillips}, {Bame}, {Gary}, {Gosling}, {Scime}, \& {Forsyth}}]{Phillips1996}
{Phillips}, J.~L., {Bame}, S.~J., {Gary}, S.~P., {et~al.} 1995, \bibinfo{title}{{Radial and Meridional Trends in Solar Wind Thermal Electron Temperature and Anisotropy: ULYSSES},} \ssr, 72, 109, \dodoi{10.1007/BF00768763}

\bibitem[{V. {Pizzo}(1978){Pizzo}}]{Pizzo1978}
{Pizzo}, V. 1978, \bibinfo{title}{{A three-dimensional model of corotating streams in the solar wind. 1. Theoretical foundations},} \jgr, 83, 5563, \dodoi{10.1029/JA083iA12p05563}

\bibitem[{P. {Planet} {et~al.}(2024){Planet}, {Badman}, {Stevens}, {Rivera}, {Panuco}, {Niembro Hernandez}, {Paulson}, {Das}, {Terres}, {Fraschetti}, \& {Bale}}]{Planet2024_AGU}
{Planet}, P., {Badman}, S.~T., {Stevens}, M.~L., {et~al.} 2024, in AGU Fall Meeting Abstracts, Vol. 2024, AGU Fall Meeting Abstracts, SH31F--2675

\bibitem[{M. {Pulupa} {et~al.}(2017){Pulupa}, {Bale}, {Bonnell}, {Bowen}, {Carruth}, {Goetz}, {Gordon}, {Harvey}, {Maksimovic}, {Mart{\'\i}nez-Oliveros}, {Moncuquet}, {Saint-Hilaire}, {Seitz}, \& {Sundkvist}}]{Pulupa2017}
{Pulupa}, M., {Bale}, S.~D., {Bonnell}, J.~W., {et~al.} 2017, \bibinfo{title}{{The Solar Probe Plus Radio Frequency Spectrometer: Measurement requirements, analog design, and digital signal processing},} Journal of Geophysical Research (Space Physics), 122, 2836, \dodoi{10.1002/2016JA023345}

\bibitem[{N.~E. {Raouafi} {et~al.}(2023){Raouafi}, {Matteini}, {Squire}, {Badman}, {Velli}, {Klein}, {Chen}, {Matthaeus}, {Szabo}, {Linton}, {Allen}, {Szalay}, {Bruno}, {Decker}, {Akhavan-Tafti}, {Agapitov}, {Bale}, {Bandyopadhyay}, {Battams}, {Ber{\v{c}}i{\v{c}}}, {Bourouaine}, {Bowen}, {Cattell}, {Chandran}, {Chhiber}, {Cohen}, {D'Amicis}, {Giacalone}, {Hess}, {Howard}, {Horbury}, {Jagarlamudi}, {Joyce}, {Kasper}, {Kinnison}, {Laker}, {Liewer}, {Malaspina}, {Mann}, {McComas}, {Niembro-Hernandez}, {Nieves-Chinchilla}, {Panasenco}, {Pokorn{\'y}}, {Pusack}, {Pulupa}, {Perez}, {Riley}, {Rouillard}, {Shi}, {Stenborg}, {Tenerani}, {Verniero}, {Viall}, {Vourlidas}, {Wood}, {Woodham}, \& {Woolley}}]{Raouafi2023}
{Raouafi}, N.~E., {Matteini}, L., {Squire}, J., {et~al.} 2023, \bibinfo{title}{{Parker Solar Probe: Four Years of Discoveries at Solar Cycle Minimum},} \ssr, 219, 8, \dodoi{10.1007/s11214-023-00952-4}

\bibitem[{Y. {Rivera} {et~al.}(2025b){Rivera}, {Finley}, {Badman}, {Klein}, {Stevens}, {Paulson}, {Niembro Hernandez}, {Das}, {Terres}, {Fraschetti}, {Halekas}, {Livi}, {Larson}, {Rahmati}, {Whittlesey}, {Velli}, {Huang}, {Romeo}, {Owen}, {Horbury}, {Ervin}, {Bale}, \& {Raouafi}}]{Rivera2025_AGU}
{Rivera}, Y., {Finley}, A., {Badman}, S.~T., {et~al.} 2025b, in AGU Fall Meeting Abstracts, Vol. 2025, AGU Fall Meeting Abstracts, SH32A--03

\bibitem[{Y.~J. {Rivera}(et al., in prep.){Rivera}}]{Rivera2026_inprep}
{Rivera}, Y.~J. et al., in prep., \bibinfo{title}{A Multi-Species Energy Budget of Polar Coronal Hole-like Solar Wind from its Sub-Alfvenic state to 1au,} in prep.

\bibitem[{Y.~J. {Rivera} \& S.~T. {Badman}(2025){Rivera} \& {Badman}}]{Rivera2025_IAU}
{Rivera}, Y.~J., \& {Badman}, S.~T. 2025, \bibinfo{title}{{An assessment of observational coverage and gaps for robust Sun to heliosphere integrated science},} arXiv e-prints, arXiv:2502.06036, \dodoi{10.48550/arXiv.2502.06036}

\bibitem[{Y.~J. {Rivera} {et~al.}(2024){Rivera}, {Badman}, {Stevens}, {Verniero}, {Stawarz}, {Shi}, {Raines}, {Paulson}, {Owen}, {Niembro}, {Louarn}, {Livi}, {Lepri}, {Kasper}, {Horbury}, {Halekas}, {Dewey}, {De Marco}, \& {Bale}}]{Rivera2024}
{Rivera}, Y.~J., {Badman}, S.~T., {Stevens}, M.~L., {et~al.} 2024, \bibinfo{title}{{In situ observations of large-amplitude Alfv{\'e}n waves heating and accelerating the solar wind},} Science, 385, 962, \dodoi{10.1126/science.adk6953}

\bibitem[{Y.~J. {Rivera} {et~al.}(2025a){Rivera}, {Badman}, {Verniero}, {Varesano}, {Stevens}, {Stawarz}, {Reeves}, {Raines}, {Raymond}, {Owen}, {Livi}, {Lepri}, {Landi}, {Halekas}, {Ervin}, {Dewey}, {De Marco}, {D'Amicis}, {Dakeyo}, {Bale}, \& {Alterman}}]{Rivera2025}
{Rivera}, Y.~J., {Badman}, S.~T., {Verniero}, J.~L., {et~al.} 2025a, \bibinfo{title}{{Differentiating the Acceleration Mechanisms in the Slow and Alfv{\'e}nic Slow Solar Wind},} \apj, 980, 70, \dodoi{10.3847/1538-4357/ada699}

\bibitem[{O.~M. {Romeo} {et~al.}(2023){Romeo}, {Braga}, {Badman}, {Larson}, {Stevens}, {Huang}, {Phan}, {Rahmati}, {Livi}, {Alnussirat}, {Whittlesey}, {Szabo}, {Klein}, {Niembro-Hernandez}, {Paulson}, {Verniero}, {Lario}, {Raouafi}, {Ervin}, {Kasper}, {Pulupa}, {Bale}, \& {Linton}}]{Romeo2023}
{Romeo}, O.~M., {Braga}, C.~R., {Badman}, S.~T., {et~al.} 2023, \bibinfo{title}{{Near-Sun In Situ and Remote-sensing Observations of a Coronal Mass Ejection and its Effect on the Heliospheric Current Sheet},} \apj, 954, 168, \dodoi{10.3847/1538-4357/ace62e}

\bibitem[{T. {Rotter} {et~al.}(2012){Rotter}, {Veronig}, {Temmer}, \& {Vr{\v{s}}nak}}]{Rotter2012}
{Rotter}, T., {Veronig}, A.~M., {Temmer}, M., \& {Vr{\v{s}}nak}, B. 2012, \bibinfo{title}{{Relation Between Coronal Hole Areas on the Sun and the Solar Wind Parameters at 1 AU},} \solphys, 281, 793, \dodoi{10.1007/s11207-012-0101-y}

\bibitem[{D. {Ruffolo} {et~al.}(2020){Ruffolo}, {Matthaeus}, {Chhiber}, {Usmanov}, {Yang}, {Bandyopadhyay}, {Parashar}, {Goldstein}, {DeForest}, {Wan}, {Chasapis}, {Maruca}, {Velli}, \& {Kasper}}]{Ruffolo2020}
{Ruffolo}, D., {Matthaeus}, W.~H., {Chhiber}, R., {et~al.} 2020, \bibinfo{title}{{Shear-driven Transition to Isotropically Turbulent Solar Wind Outside the Alfv{\'e}n Critical Zone},} \apj, 902, 94, \dodoi{10.3847/1538-4357/abb594}

\bibitem[{E. {Sanchez-Diaz} {et~al.}(2016){Sanchez-Diaz}, {Rouillard}, {Lavraud}, {Segura}, {Tao}, {Pinto}, {Sheeley}, \& {Plotnikov}}]{Sanchez-Diaz2016}
{Sanchez-Diaz}, E., {Rouillard}, A.~P., {Lavraud}, B., {et~al.} 2016, \bibinfo{title}{{The very slow solar wind: Properties, origin and variability},} Journal of Geophysical Research (Space Physics), 121, 2830, \dodoi{10.1002/2016JA022433}

\bibitem[{K.~H. {Schatten} {et~al.}(1969){Schatten}, {Wilcox}, \& {Ness}}]{Schatten1969_pfss}
{Schatten}, K.~H., {Wilcox}, J.~M., \& {Ness}, N.~F. 1969, \bibinfo{title}{{A model of interplanetary and coronal magnetic fields},} \solphys, 6, 442, \dodoi{10.1007/BF00146478}

\bibitem[{C. {Shi} {et~al.}(2022{\natexlab{a}}){Shi}, {Velli}, {Bale}, {R{\'e}ville}, {Maksimovi{\'c}}, \& {Dakeyo}}]{shi2022_polyt}
{Shi}, C., {Velli}, M., {Bale}, S.~D., {et~al.} 2022{\natexlab{a}}, \bibinfo{title}{{Acceleration of polytropic solar wind: Parker Solar Probe observation and one-dimensional model},} Physics of Plasmas, 29, 122901, \dodoi{10.1063/5.0124703}

\bibitem[{C. {Shi} {et~al.}(2022{\natexlab{b}}){Shi}, {Panasenco}, {Velli}, {Tenerani}, {Verniero}, {Sioulas}, {Huang}, {Brosius}, {Bale}, {Klein}, {Kasper}, {de Wit}, {Goetz}, {Harvey}, {MacDowall}, {Malaspina}, {Pulupa}, {Larson}, {Livi}, {Case}, \& {Stevens}}]{Shi2022}
{Shi}, C., {Panasenco}, O., {Velli}, M., {et~al.} 2022{\natexlab{b}}, \bibinfo{title}{{Patches of Magnetic Switchbacks and Their Origins},} \apj, 934, 152, \dodoi{10.3847/1538-4357/ac7c11}

\bibitem[{N. {Sioulas} {et~al.}(2026){Sioulas}, {Velli}, {Shi}, {Matteini}, {Bowen}, {Mallet}, {Larosa}, {Tenerani}, \& {Horbury}}]{Sioulas26}
{Sioulas}, N., {Velli}, M., {Shi}, C., {et~al.} 2026, \bibinfo{title}{{Generation and Expansion-Driven Growth of Switchbacks in the Outer Solar Corona and Solar Wind},} arXiv e-prints, arXiv:2602.03724, \dodoi{10.48550/arXiv.2602.03724}

\bibitem[{I.~V. {Sokolov} {et~al.}(2013){Sokolov}, {van der Holst}, {Oran}, {Downs}, {Roussev}, {Jin}, {Manchester}, {Evans}, \& {Gombosi}}]{Sokolov2013}
{Sokolov}, I.~V., {van der Holst}, B., {Oran}, R., {et~al.} 2013, \bibinfo{title}{{Magnetohydrodynamic Waves and Coronal Heating: Unifying Empirical and MHD Turbulence Models},} \apj, 764, 23, \dodoi{10.1088/0004-637X/764/1/23}

\bibitem[{D. {Stansby} {et~al.}(2020){Stansby}, {Yeates}, \& {Badman}}]{Stansby2020}
{Stansby}, D., {Yeates}, A., \& {Badman}, S. 2020, \bibinfo{title}{{pfsspy: A Python package for potential field source surface modelling},} The Journal of Open Source Software, 5, 2732, \dodoi{10.21105/joss.02732}

\bibitem[{{\v{S}}. {\v{S}}tver{\'a}k {et~al.}(2025){\v{S}}tver{\'a}k, Her{\v{c}}{\'i}k, Nicolaou, Hellinger, Pop{\v{d}}akunik, Khotyaintsev, Kataria, Owen, \& Maksimovic}]{stverak2025}
{\v{S}}tver{\'a}k, {\v{S}}., Her{\v{c}}{\'i}k, D., Nicolaou, G., {et~al.} 2025, \bibinfo{title}{Effects of cold electron emissions on thermal plasma measurements on board {Solar Orbiter} spacecraft,} Astronomy \& Astrophysics, 693, A185, \dodoi{10.1051/0004-6361/202452030}

\bibitem[{ {SunPy Community} {et~al.}(2020){SunPy Community}, {Barnes}, {Bobra}, {Christe}, {Freij}, {Hayes}, {Ireland}, {Mumford}, {Perez-Suarez}, {Ryan}, {Shih}, {Chanda}, {Glogowski}, {Hewett}, {Hughitt}, {Hill}, {Hiware}, {Inglis}, {Kirk}, {Konge}, {Mason}, {Maloney}, {Murray}, {Panda}, {Park}, {Pereira}, {Reardon}, {Savage}, {Sip{\H{o}}cz}, {Stansby}, {Jain}, {Taylor}, {Yadav}, {Rajul}, \& {Dang}}]{Sunpy2020}
{SunPy Community}, {Barnes}, W.~T., {Bobra}, M.~G., {et~al.} 2020, \bibinfo{title}{{The SunPy Project: Open Source Development and Status of the Version 1.0 Core Package},} \apj, 890, 68, \dodoi{10.3847/1538-4357/ab4f7a}

\bibitem[{A.~V. {Usmanov} {et~al.}(2025){Usmanov}, {Chhiber}, {Matthaeus}, {Roy}, \& {Goldstein}}]{Usmanov2025}
{Usmanov}, A.~V., {Chhiber}, R., {Matthaeus}, W.~H., {Roy}, S., \& {Goldstein}, M.~L. 2025, \bibinfo{title}{{A Unified Three-dimensional Magnetohydrodynamic Model of the Solar Corona, Solar Wind, and Global Heliosphere with Turbulence Transport},} \apj, 993, 87, \dodoi{10.3847/1538-4357/ae019c}

\bibitem[{A. {Verdini} \& M. {Velli}(2007){Verdini} \& {Velli}}]{Verdini2007}
{Verdini}, A., \& {Velli}, M. 2007, \bibinfo{title}{{Alfv{\'e}n Waves and Turbulence in the Solar Atmosphere and Solar Wind},} \apj, 662, 669, \dodoi{10.1086/510710}

\bibitem[{A. {Verdini} {et~al.}(2010){Verdini}, {Velli}, {Matthaeus}, {Oughton}, \& {Dmitruk}}]{Verdini2010}
{Verdini}, A., {Velli}, M., {Matthaeus}, W.~H., {Oughton}, S., \& {Dmitruk}, P. 2010, \bibinfo{title}{{A Turbulence-Driven Model for Heating and Acceleration of the Fast Wind in Coronal Holes},} \apjl, 708, L116, \dodoi{10.1088/2041-8205/708/2/L116}

\bibitem[{R. {von Steiger} \& T.~H. {Zurbuchen}(2011){von Steiger} \& {Zurbuchen}}]{vonSteiger2011}
{von Steiger}, R., \& {Zurbuchen}, T.~H. 2011, \bibinfo{title}{{Polar coronal holes during the past solar cycle: Ulysses observations},} Journal of Geophysical Research (Space Physics), 116, A01105, \dodoi{10.1029/2010JA015835}

\bibitem[{R. {von Steiger} {et~al.}(2000){von Steiger}, {Schwadron}, {Fisk}, {Geiss}, {Gloeckler}, {Hefti}, {Wilken}, {Wimmer-Schweingruber}, \& {Zurbuchen}}]{vonsteiger2000}
{von Steiger}, R., {Schwadron}, N.~A., {Fisk}, L.~A., {et~al.} 2000, \bibinfo{title}{{Composition of quasi-stationary solar wind flows from Ulysses/Solar Wind Ion Composition Spectrometer},} \jgr, 105, 27217, \dodoi{10.1029/1999JA000358}

\bibitem[{Y.-M. {Wang} \& N.~R. {Sheeley}(1990){Wang} \& {Sheeley}}]{Wang1990}
{Wang}, Y.-M., \& {Sheeley}, Jr., N.~R. 1990, \bibinfo{title}{{Solar Wind Speed and Coronal Flux-Tube Expansion},} \apj, 355, 726, \dodoi{10.1086/168805}

\bibitem[{P.~L. {Whittlesey} {et~al.}(2020){Whittlesey}, {Larson}, {Kasper}, {Halekas}, {Abatcha}, {Abiad}, {Berthomier}, {Case}, {Chen}, {Curtis}, {Dalton}, {Klein}, {Korreck}, {Livi}, {Ludlam}, {Marckwordt}, {Rahmati}, {Robinson}, {Slagle}, {Stevens}, {Tiu}, \& {Verniero}}]{Whittlesey2020}
{Whittlesey}, P.~L., {Larson}, D.~E., {Kasper}, J.~C., {et~al.} 2020, \bibinfo{title}{{The Solar Probe ANalyzers{\textemdash}Electrons on the Parker Solar Probe},} \apjs, 246, 74, \dodoi{10.3847/1538-4365/ab7370}

\bibitem[{L.~B. {Wilson} {et~al.}(2023){Wilson}, {Stevens}, {Kasper}, {Klein}, {Maruca}, {Bale}, {Bowen}, {Pulupa}, \& {Salem}}]{Wilson2023}
{Wilson}, L.~B., {Stevens}, M.~L., {Kasper}, J.~C., {et~al.} 2023, \bibinfo{title}{{Erratum: ``The Statistical Properties of Solar Wind Temperature Parameters Near 1 au'' (2018, ApJS, 236, 41)},} \apjs, 269, 62, \dodoi{10.3847/1538-4365/ad07de}

\bibitem[{L.~B. {Wilson} {et~al.}(2026){Wilson}, {Bale}, {Stevens}, {Maruca}, {Klein}, {Martinovi{\'c}}, \& {TenBarge}}]{Wilson2026}
{Wilson}, III, L.~B., {Bale}, S.~D., {Stevens}, M.~L., {et~al.} 2026, \bibinfo{title}{{Electron Velocity Moments in the Solar Wind. II. Statistical Properties},} \apj, 1002, 122, \dodoi{10.3847/1538-4357/ae5d4a}

\bibitem[{L.~B. {Wilson III} {et~al.}(2023{\natexlab{a}}){Wilson III}, {Salem}, \& {Bonnell}}]{wilsoniii23a}
{Wilson III}, L.~B., {Salem}, C.~S., \& {Bonnell}, J.~W. 2023{\natexlab{a}}, \bibinfo{title}{{Spacecraft floating potential measurements for the \emph{Wind} spacecraft},} Astrophys. J. Suppl., 269, 10, \dodoi{10.3847/1538-4365/ad0633}

\bibitem[{L.~B. {Wilson III} {et~al.}(2023{\natexlab{b}}){Wilson III}, {Salem}, \& {Bonnell}}]{wilsoniii23b}
{Wilson III}, L.~B., {Salem}, C.~S., \& {Bonnell}, J.~W. 2023{\natexlab{b}}, \bibinfo{title}{\emph{Wind} spacecraft floating potential measurements,}, 1.0 Zenodo, \dodoi{10.5281/zenodo.8364797}

\bibitem[{L.~B. {Wilson III} {et~al.}(2021){Wilson III}, {Brosius}, {Gopalswamy}, {Nieves-Chinchilla}, {Szabo}, {Hurley}, {Phan}, {Kasper}, {Lugaz}, {Richardson}, {Chen}, {Verscharen}, {Wicks}, \& {TenBarge}}]{wilsoniii21a}
{Wilson III}, L.~B., {Brosius}, A.~L., {Gopalswamy}, N., {et~al.} 2021, \bibinfo{title}{{A Quarter Century of \emph{Wind} Spacecraft Discoveries},} Rev. Geophys., 59, e2020RG000714, \dodoi{10.1029/2020RG000714}

\bibitem[{T. {Woolley} {et~al.}(2020){Woolley}, {Matteini}, {Horbury}, {Bale}, {Woodham}, {Laker}, {Alterman}, {Bonnell}, {Case}, {Kasper}, {Klein}, {Martinovi{\'c}}, \& {Stevens}}]{Woolley2020}
{Woolley}, T., {Matteini}, L., {Horbury}, T.~S., {et~al.} 2020, \bibinfo{title}{{Proton core behaviour inside magnetic field switchbacks},} \mnras, 498, 5524, \dodoi{10.1093/mnras/staa2770}

\bibitem[{X. Wu {et~al.}(2026)Wu, Owen, Coburn, Nicolaou, Verscharen, Liu, Ioannou, Ran, Rivera, \& Yardley}]{wu2026}
Wu, X., Owen, C.~J., Coburn, J., {et~al.} 2026, \bibinfo{title}{Correlation between Electron Temperature and Ion Charge-state Ratios in the Solar Wind at {$\sim$}0.5 au,} The Astrophysical Journal, 1000, 13, \dodoi{10.3847/1538-4357/ae3c7b}

\bibitem[{L. Zhao \& E. Landi(2014)Zhao \& Landi}]{Zhao_2014}
Zhao, L., \& Landi, E. 2014, \bibinfo{title}{POLAR AND EQUATORIAL CORONAL HOLE WINDS AT SOLAR MINIMA: FROM THE HELIOSPHERE TO THE INNER CORONA,} The Astrophysical Journal, 781, 110, \dodoi{10.1088/0004-637X/781/2/110}

\bibitem[{A.~N. {Zhukov} {et~al.}(2026){Zhukov}, {Patel}, {Debrabandere}, {Dolla}, {Mierla}, {Shestov}, {Bourgoignie}, {Jean}, {Nicula}, {Talpeanu}, {Zontou}, {Fineschi}, {Gun{\'a}r}, {Lamy}, {Peter}, {Rudawy}, {Tsinganos}, {Abbo}, {Aime}, {Andretta}, {Auch{\`e}re}, {Berghmans}, {Be{\textcommabelow s}liu-Ionescu}, {Gibson}, {Giordano}, {Heinzel}, {Inhester}, {Magdaleni{\'c}}, {Marqu{\'e}}, {Rodriguez}, {Ste{\'s}licki}, {Zangrilli}, {Galano}, {Rougeot}, {Versluys}, \& {Thizy}}]{Zhukov2026}
{Zhukov}, A.~N., {Patel}, B.~D., {Debrabandere}, A., {et~al.} 2026, \bibinfo{title}{{Ubiquitous Small-scale Dynamics in the Slow Solar Wind Formation Region Observed by Proba-3/ASPIICS},} \apjl, 999, L41, \dodoi{10.3847/2041-8213/ae469b}

\end{thebibliography}



\end{document}